\magnification = 1200    
\baselineskip =0.6 true cm
\input lanlmac.tex

\input epsf

\ifx\epsfbox\UnDeFiNeD\message{(NO epsf.tex, FIGURES WILL BE IGNORED)}
\def\figin#1{\vskip2in}
\else\message{(FIGURES WILL BE INCLUDED)}\def\figin#1{#1}\fi
\def\tfig#1{{\xdef#1{Fig.\thinspace\the\figno}}
Fig.\thinspace\the\figno \global\advance\figno by1}
%


%
%
%

%
%
%
%
\font\tit=cmr10 scaled \magstep2
\def\chapter#1{\vskip 1cm$\bf\hbox{#1}$}     
     
\def\underrel#1#2{\mathrel{\mathop{\kern0pt #1}\limits_{#2}}}
\def\:{\quad}
\def\;{\qquad}
\def\hchi{\raise1pt\hbox{$\chi$}}
     
\centerline{\tit One-Dimensional String Theory with Vortices }
\centerline{\tit as  Upside-Down Matrix Oscillator}
\vskip 1cm 
\centerline{ Dmitri Boulatov}
\centerline{\it Institute of Theoretical and
Experimental Physics}
\centerline{\it B.Cheremushkinskaya 25, 117259 Moscow, RUSSIA}
 \centerline{and}
\centerline{Vladimir Kazakov}
\centerline{\it Laboratoire de
 Physique Th\'{e}orique de l'Ecole Normale  Sup\'{e}rieure, }
\centerline{\it
 24, rue Lhomond, 75231 Paris Cedex 05, FRANCE
\footnote\ddag{\rm Unit\'e de Recherche du
Centre National de la Recherche Scientifique et de l'Ecole Normale
Sup\'erieure et \`a l'Universit\'e de Paris-Sud.}}
\vskip 1cm
\centerline{\tit Abstract}

We study matrix quantum mechanics at a finite temperature equivalent
to one dimensional compactified string theory with vortex (winding)
excitations.  It is explicitly demonstrated that the states
transforming under non-trivial U(N) representations describe various
configurations vortices and anti-vortices.  For example, for the
adjoint representation the Feynman graphs (representing discretized
world-sheets) contain two faces with the boundaries wrapping around
the compactified target space which is equivalent to a
vortex-anti-vortex pair.  A technique is developed to calculate
partition functions in a given representation for the standard matrix
oscillator.  It enables us to obtain the partition function in the
presence of a vortex-anti-vortex pair in the double scaling limit
using an analytical continuation to the upside-down oscillator. The
Berezinski-Kosterlitz-Thouless phase transition occurs in a similar
way and at the same temperature as in the flat 2D space.  A possible
generalization of our technique to any dimension of the embedding
space is discussed.   

\line{\hfill}
\leftline{LPTENS 91/24}     
\line{KUNS 1094 HE(TH) 91/14 \hfill     August 1991}

\vfill
\eject


\chapter{1. Introduction}

Matrix quantum mechanics represents an important analytical tool for
 the investigation of one dimensional bosonic string theory (see
 references in review [1]).  Matrix formulation of this string theory
 is based on the correspondence between the feynman diagrammatic
 expansion of the matrix functional integral and the sum over all
 possible discretized (say, triangulated) curved two-dimensional
 manifolds embedded into one dimensional target space (time).

 It was claimed in ref. [2] that the singlet sector of the quantum
mechanical matrix model (trivial representation of U(N)) compactified
on a time circle of the length $\beta$ represents the vortex free
sector of one dimensional bosonic string theory.  This claim was
supported by a remarkable self-duality with respect to the change
${\beta } \rightarrow {\alpha \prime /\beta }$ expected from the
continuum string field theory.

Non-singlet states are of special interest in the $1D$ matrix theory.
It was conjectured in [2-3] that they correspond to the
vortex-anti-vortex excitations (winding modes) of compactified $1D$
bosonic string with the target space coordinate living on a circle of a
finite length $\beta$. In the statistical mechanics language, they
describe the classical $XY$-model on a random (dynamical) 2D manifold.
     
In ref. [3] the simplest non-singlet case, (the adjoint
representation for ``angular'' variables), was investigated.  Using
the results of the paper [4] where the planar $(N \rightarrow \infty
)$ approximation for the adjoint hamiltonian was developed, the
authors of ref. [3] have demonstrated that there exists a big gap
separating ground states for the singlet and the adjoint
representations:
$$
\eqalign{
E_{adj} - E_{sing}  \underrel{\sim} {\Delta \to 0}
{\beta  \over 2\pi } \vert \log \Delta \vert
}
\eqno{(1.1)}
$$
where $\Delta  = \lambda_c  - \lambda$ is the deviation of the
 cosmological constant from its critical value.
     
It was also argued in ref. [3] that this result can be qualitatively
reproduced from the continuum Liouville theory assuming that one
includes a vortex-anti-vortex pair into the system.  So, the adjoint
representation was conjectured to describe a vortex-anti-vortex pair
in the presence of 2D quantum gravity.  Higher representations were
associated with bigger numbers of vortices.  On the base of these
conjectures the authors of ref. [3] gave an estimate of the position,
$\beta _c$, of the Berezinski-Kostelitz-Thouless phase transition
arguing that $\beta _{KT}$ is finite.  The phase $\beta >
\beta _{KT}$ (low vortex density) describes $c = 1$ system  and, for
$\beta < \beta _{KT}$, $c=0$ (vortices destroy long order correlations).

In this paper we shall use a different technique but the result (1.1)
will be confirmed and even generalized. Namely, we shall calculate the
double scaling limit of the partition function for the adjoint
representation providing the partition function of vortex-anti-vortex
pair interacting with the 2D quantum gravity to all genera.  We shall
explicitly demonstrate by direct arguments that for the matrix quantum
mechanics in the adjoint representation describes the
``vortex-anti-vortex'' sector of the $XY$-model (sometimes called as
the model of planar rotators) interacting with 2D quantum gravity. As
in the $XY$-model on a regular lattice, knowing the partition function
in this sector one can determine $\beta_{KT}$ precisely. Our estimate
$\beta_{KT} = 4\pi$ coincides with the one given in ref. [3].

We will propose here a mathematical setup, different from [3], for the
analysis of nonsinglet sectors in the matrix quantum mechanics.
Although we can solve a number of matrix models exactly we usually
need from the solutions only a small piece of information about the
continuum (thermodynamic) limit of lattice manifolds dominated by huge
Feynman graphs. In this respect it seems to be very desirable to have
in hands some formalism which has an advantage of the matrix approach
to be manifestly invariant (i.e. not using the notion of a coordinate
system and non-invariant objects like a metric) but avoiding the
intermediate lattice stage usual for the matrix models and dealing
directly with the continuum theory.
     
In the matrix language one usually diagonalizes the matrices and
tries to formulate a model in terms of their eigenvalues 
integrating explicitly, if it appears to be possible, over the
``angular'' degrees of freedom belonging to a group (U(N) in the case
of $N \times N$ Hermite-an matrices).  For many interesting models of
2d gravity with matter fields having the central charge $c
\leq 1$, this integration can be performed exactly.  In the resulting
effective action for the eigenvalues, not the entire potential of the
original matrix model is important for the continuum (planar or double
scaling) limit but only the vicinity of a singularity of the effective
action.  In the most generic case, this is a quadratic top of the
potential.  The behavior of the eigenvalues near the top usually
defines the essential, long distance phenomena and everything else in
the shape of the potential is responsible only for microscopic
details, {\sl i.e.}  cut-off, area definition etc.  This brings to the
mind the idea that the matrix models can be formulated in a simplified
way where only the essential part of the matrix potential will be
taken into account. One can hope that, in the continuum limit, the
whole model might appear to be almost ({\sl i.e.} up to a cut-off)
Gaussian and, hence, exactly solvable even though the potential is
unstable.
     
One of the most transparent examples realizing this mechanism is the
matrix quantum mechanics [5], which was shown to correspond to the
bosonic string field theory in one-dimensional target space [6] and
was solved later in the double scaling limit [7-10].  This theory can
be described by matrix quantum mechanics with an upside-down
oscillatorial potential.  The system can be stabilized by putting
infinite walls on cut-off distances $\pm \Lambda $ from the top of the
potential for every eigenvalue of the matrix field.
     
In this case it is natural to expect that interesting physical
quantities can be obtained by an analytical continuation of
corresponding quantities for the standard matrix oscillator with the
stable quadratic potential, {\sl i.e.} by the change $\omega
\rightarrow i\omega $ (where $\omega $ is the frequency of the
oscillator).
     
Life appears to be more complicated: the standard oscillator has
more symmetry then the upside-down one and it does not know about any
cut-off.  Usually to establish a correspondence between them the
direct analytical continuation is not possible and should be completed
by a guess about the cut-off dependence.  One of the recent results of
this correspondence was given in ref. [11] for the singlet sector
Green function where the corresponding planar result of ref. [12] was
generalized to the double scaling limit.

In the present paper we give another example of such correspondence.
Our goal will be to calculate partition functions for some non-singlet
states in the $1D$ model in the double scaling limit using partition
functions for the ordinary matrix oscillator. We argue that this
correspondence works at least for a sector of the model, given by the
adjoint representation of U$(N)$ and (as we show explicitly)
describing single vortex-anti-vortex configuration. At the end we
discuss a possible generalization of this approach to a $D+1$
dimensional matrix model which might give in the future new analytical
tools into our hands.

The paper is organized as follows.

In section 2, we formulate the matrix model with the compactified time
and discuss its correspondence with $1D$ string theory and the
$XY$-model.

In section 3, we give a derivation of Hamiltonians in non-singlet
sectors in terms of the eigenvalues of the matrix field.  The double
scaling is considered for which the problem can be reformulated as the
upside-down matrix oscillator.

In section 4, a $U(N)$ twisted partition function is introduced which
 allows us to expand the oscillatorial partition function into a sum
 of partition functions of states in given representations of
 $U(N)$. In this section we also show that the adjoint sector of the
 $1D$ matrix model corresponds to the insertion of one
 vortex-anti-vortex pair on the string world sheet.

In section 5, the analytical continuation of the adjoint partition
function is suggested and, in section 6, physical consequences of the
obtained results are considered.

In section 7, a possible generalization of our approach to the case of
$D+1$ string theory is sketched out using the idea of correspondence
between Gaussian matrix models with the stable and the upside-down
quadratic potentials.

The section 8 is devoted to  conclusion.

Three appendices contain technical details  concerning 
applications of the group theory methods to our model.
     
\chapter{2. Matrix quantum mechanics in periodic time}
  {\bf as 1D compactified string theory}

\def\calZ{{\cal Z}}
     
The matrix model under consideration can be defined by the following
functional integral for the partition function:
$$
\eqalignno{
 \calZ_N(\beta , \lambda ) = & \int_
 {\varphi (0) = \varphi (\beta )}
 {\cal D}^{N^2}\varphi (t)  \exp - Ntr
 \int ^\beta _0 dt \big[ {1 \over 2} {\dot \varphi}^2 + V(\varphi)\big]
 & (2.1)
}
$$
where $\varphi (t)$ is a  hermitian-matrix-valued periodic $1D$ field :
$\varphi _{ij} (t) = \overline \varphi _{ji} (t)$; 
 $\varphi_{ij} (0) = \varphi_{ij} (\beta )$ $i,j=1,\ldots,N$.
The potential can be chosen, for example, in the form
$$
\eqalign{
V(\varphi ) = {1 \over 2} \varphi ^2 - {\lambda  \over 3} \varphi ^3,
}
\eqno{(2.2)}
$$
which corresponds to triangulations of the string world sheet.
     
The standard arguments leading to the string picture interpretation of
eq. (2.1) are the following [3]. Let us expand $log \calZ_N(\beta , \lambda )$
in the coupling $\lambda $ using diagrammatic Feynman rules.
The corresponding free energy
$f_N (\beta , \lambda ) = {1 \over N^2} \log  \calZ _N(\beta , \lambda )$
has the form:
$$
\eqalignno{
f_N(\beta , \lambda ) & =
       \sum^\infty _{g=0} N^{-2g} \sum ^\infty _{V=0} \lambda ^V
       \sum_{G^{(V)}_g} \int ^\beta _0 \cdots \int ^\beta _0 dt_1
       \cdots dt_V \times \cr
& \prod _{<ij>} \sum^\infty _{m_{ij}=-\infty } e^{-\mid t_i - t_j +
 \beta m_{ij}
\mid} & (2.3)
}
$$
where $g$ is the genus of a $\varphi ^3$-Feynman graph $G^{(V)}_g$;
$V$ is the number of vertices, $\prod_{< ij >}$ is the product over all links of the
graph, $<ij> \in G^{(n)}_g$, and the periodic propagator in our case is
$$
\eqalign{
D(t_i - t_j) = \sum^\infty _{m=-\infty } e^{-\mid t_i - t_j + m\beta
 \mid}
}
\eqno{(2.4)}
$$
where $t_i$'s are attached to vertices of a graph. 
The constant $\lambda $ plays the role of a bare cosmological constant
and $1/N$ is the (topological) string coupling constant.
For sufficiently large $\beta $ we expect that only the term with
$m=0$ in eq. (2.4) will survive and we will come to the well investigated
case of the $1D$ matrix model on the infinite time interval [1].
 In this case time
coordinates $t_i, i= 1,2 \ldots V$, will correspond to the target
space coordinates of the string world sheet represented by a
$\varphi ^3$-graph $G^{(V)}_g$.
The corresponding action is of the form
$$
\eqalign{
S = \sum_{<ij> \in G^{(V)}_g} \mid t_i - t_j \mid
}
\eqno{(2.5)}
$$
which differs from the discretized Polyakov string
action :
$$
\eqalign{
S_p = \sum _{<ij> \in G^{(V)}_g} (t_i - t_j)^2
}
\eqno{(2.6)}
$$
but is proven to be in the same universality class due to the
superconvergency of underlying Feynman diagrams.
     
Eq. (2.3) describes the $XY$-model coupled to $2d$ gravity and we
expect the appearance of all phenomena related to the vortex
configurations of the $t$-field.  The analysis of the vortex
kinematics in the model was done in ref. [2].  Let us briefly mention
it here for completeness.  Following the classical papers [13],[14],
[15] we can view $m_{ij}$ in eq.(2.3) as an abelian gauge field
defined on links of a $\varphi ^3$-graph.  The sum along a
non-self-intersecting loop $L$ on the graph
$$
\eqalign{
M_L = \sum_{<ij> \in L } m_{ij}
}
\eqno{(2.7)}
$$
gives the integral charge of the vortices enveloped by the contour
$L$.  In the case when $L$ is just the boundary of a face of the
graph, $M_L$ can be considered as the elementary field strength or the
vortex number in this face.  The duality transformation corresponding
to the Fourier transform of the original propagators
$$
\eqalign{
\sum ^\infty _{m=-\infty } e^{-\mid t_i - t_j + \beta m\mid} =
{2 \over \beta } \sum^\infty _{k_{ij}=-\infty } e^{i{2\pi  \over \beta
 }k_{ij}(t_i - t_j)}
 {1 \over 1 + \big({2\pi \over \beta }k_{ij} \big)^2}
}
\eqno{(2.8)}
$$
brings eq. (2.3) to the form where $M_L$ are new dynamical variables.
If we insert eq. (2.8) into eq. (2.3) and integrate over $t_i$'s, we
obtain the constraint at every vertex $i$, which is just the ordinary
condition of the momentum conservation,
$$
\eqalign{
 k_{ij_1} + k_{ij_2} + k_{ij_3} = 0
}
\eqno{(2.9)}
$$
It means that (excluding the zero modes) we have got $E-V+1 = F-1+2g$
variables instead of $V-1$ (in virtue of the Euler theorem:
$F-E+V=2-2g$, where $F$, $E$ and $V$ are the numbers of faces, edges
and vertices of a graph). The usual choice of these variables is to
attach a momentum $p_I$ to each face and to define remaining $2g$
variables as momenta $l_a$ running along $2g$ independent
non-contractable loops on a graph: $L_a, a=1,\ldots,2g$.
     
The resulting dual representation for the partition function (2.3)
takes the form
$$
\eqalignno{
f_N(\beta , \lambda ) & =\bigg({\beta\over \lambda}\bigg)^2
 \sum^\infty _{g=0}
 \bigg({N\beta \over \lambda}\bigg)^{-2g}
\sum_F \bigg({\lambda^{1\over2}\over  \beta }\bigg)^F
\sum_{{\tilde G}^{(F)}_g} 
\sum _{p_1 =-\infty}^{+\infty}\ldots
\sum _{p_F =-\infty}^{+\infty}\cr
&\sum _{l_1 =-\infty}^{+\infty}\ldots
\sum _{l_{2g} =-\infty}^{+\infty}
\prod _{<IJ>}
{2\over 1 + \big({2\pi  \over \beta }\big)^2
(p_I - p_J + \sum_{a=1}^{2g}l_a {\epsilon ^a_{IJ}})^2} & (2.10)
}
$$
where $\sum_{{\tilde G}^{(F)}_g} $ is the sum over all dual graphs
(triangulations) having $F$ dual vertices and a genus $g$; $\prod
_{<IJ>}$ is the product over all dual links $<IJ> \in {\tilde
G}^{(F)}_g$; $\epsilon ^a_{IJ} = \pm 1$ when a dual edge $<IJ>$
crosses an edge belonging to a chosen in advance non-contractable
cycle $L_a$, and is zero otherwise.  The sign has to be chosen with
respect to the mutual orientation of the link $<IJ>$ and the loop
$L_a$.
     
It can be argued that one can discard the vortices in eq.(2.3) by
imposing the "pure gauge" conditions on the field $m_{ij}$ (up to a
contribution of non-contractable loops)
$$
\eqalign{
m_{ij} = \sum_{a=1}^{2g}{\tilde \epsilon}^a_{ij} \tilde l_a + m_i - m_j
}
\eqno{(2.11)}
$$
where integers $\tilde l_a$ are defined on non-contractable dual loops
$\tilde L_a$; $\tilde \epsilon_{ij}^a$ is the object dual to
$\epsilon_{IJ}^a$.  Then, eq. (2.10) will look as
$$
\eqalignno{
&f_N(\beta , \lambda )  =\bigg({\beta\over \lambda}\bigg)^2
 \sum^\infty _{g=0}
 \bigg({N\beta \over \lambda}\bigg)^{-2g}
\sum_F \bigg({\lambda^{1\over2}\over  \beta} \bigg)^F
\sum_{{\tilde G}^{(F)}_g} \cr
& \int ^\infty _{-\infty } 
\prod_{I=1}^{F} dp_I \sum ^\infty _{l_1=-\infty }\ldots
\sum _{l_{2g} =-\infty}^{+\infty}
\prod _{<IJ>} {2\over 1 +\big({2\pi  \over \beta } \big)^2
(p_I - p_J +\sum_{a=1}^{2g}l_a\epsilon ^a_{IJ})^2}
& {(2.12)}
}
$$
     
The sums over $p_I$ in eq. (2.10) are substituted by the integrals,
since, in the original expression (2.3) with the choice (2.11), we can
instead of summing over $m_i$'s spread the integrations over $t_i$'s
to the infinite interval.  Then the corresponding Fourier transforms
of propagators will be integrals rather than sums.
     
Now the original representation of $f_N(\beta , \lambda )$, eq. (2.3),
with the condition (2.11) looks very similar to its dual transform
(2.12).  The arguments of the propagators and the string coupling
constant can be matched by the simple duality transformation
$$
\eqalignno{
& {\beta \over 2\pi } \rightarrow {2\pi   \over \beta } \cr
& {1 \over N} \rightarrow {2\pi   \over \beta N} & (2.13)
}
$$
     
Two main differences are : the original $\varphi ^3$-graphs are
substituted by dual ones, and the propagators are different.  If we
believe in universality in the continuum limit , these two differences
should be insignificant on the macroscopic scale. More important is
the equivalence of the sets of variables $(t_i, \tilde l_a)$ and
$(p_I, l_a)$ in eqs. (2.3) and (2.12).  So, one can hope that, in the
double scaling limit, all answers will be invariant under the duality
transformation (2.13).  It was claimed in ref. [2] that this is really
true for the singlet sector of the model, which was identified
therefore with the vortex-free partition function (2.12). We will show
in this paper that the inclusion of the non-singlet states destroys
the self-duality, which is natural to expect in the presence of
vortices.
     
\chapter{3. Effective action for eigenvalues}
     
The first standard step to calculate the partition function (2.1) is
 the diagonalization of the matrix field:
$$
\eqalign{
\varphi _{ij}(t) = \sum^N_{k=1} \Omega ^+_{ik}(t) z_k(t)\Omega _{kj}(t)
}
\eqno{(3.1)}
$$
where $\Omega(t) \in$ U(N).
     
In terms of new variables we have
$$
\eqalign{
tr{\dot \varphi}^2 = \sum^N_{i=1} {\dot z}^2_i + \sum_{i \neq j}
(z_i - z_j)^2 \mid A_{ij} \mid ^2
}
\eqno{(3.2)}
$$
where we introduced ``the connection''
$$
\eqalign{
A_{ij}(t) = (\Omega ^+{\dot \Omega })_{ij}
}
\eqno{(3.3)}
$$
     
The periodicity condition $\varphi_{ij} (t)=\varphi_{ij}(t+\beta )$
 implies that eigenvalues, $z_k$, have to be periodic in $t$ only up
 to an arbitrary substitution $\cal P$:
$$
\eqalign{
z_k(t+\beta) = \sum_{j=1}^N {\cal P}_{kj}\ z_j(t)\ {\cal P}^{-1}_{jk}\cr
\Omega (t+\beta) = {\cal P} \Omega (t)
}
\eqno{(3.4)}
$$
which shows that the variables $A_{ij}(t)$ are not independent,
contrary to the free boundary conditions case, and that they obey the
constraint:
$$
\eqalign{
\widehat T {\rm exp}~ i\int^\beta _0 dt A(t) = {\cal P}^{-1}
}
\eqno{(3.5)}
$$
     
This makes the integral over $A(t)$ rather complicated unlike the most
of solvable models of $2d$ gravity with $c<1$.  Usually, attempts to
solve some matrix models describing gravity with matter having the
central charge $c>1$ are blocked by the difficulties to compute the
emerging integrals over the ``angular'' variables.  We hope that the
investigation of the model containing at least one nontrivial
integration might be useful for the understanding of the role of
``angular'' variables for $c>1$.

 The Dyson measure for the new
variables $z$ and $\Omega $ looks as
$$
\eqalignno{
{\cal D}\varphi (t) & = \prod _{t\in [0, \beta ]} \Delta ^2(z(t))
\prod ^N_{k=1} dz_k(t) d^{N^2} A(t) \times \cr
& \times \delta _{U(N)} (\widehat T exp\int^\beta _0 Adt,
 {\cal P}^{-1})
& (3.6)
}
$$
where $\Delta (z) = \prod _{k>m}(z_k - z_m)$ is the Vandermonde
determinant; the invariant $\delta $-function can be represented as
sum over irreducible representations
$$
\eqalign{
\delta _{U(N)}(\Omega ,  {\cal P}^{-1}) =
 \sum_R d_R \hchi _R({\cal P}\Omega )
}
\eqno{(3.7)}
$$
where $d_R$ is the dimension of the R'th representation and $\hchi
 _R(\Omega )$ is the character
$$
\eqalign{
\hchi _R(\widehat T {\rm exp}~~i \int ^\beta _0 Adt) = Tr_R
\widehat T{\rm exp}\bigg(~i \int ^\beta _0
dt \sum _{i,j}
A_{ij}\widehat \tau _{ij}^R\bigg),
}
\eqno{(3.8)}
$$
where ${\hat \tau }^R_{ij}$ is a generator of $U(N)$ in the R'th
representation.

It is well known [1] that in the case of  free boundary conditions
({\sl i.e.} $\varphi(0)$ independent of $\varphi(\beta)$) the singlet
partition function describes non-interacting fermions. Indeed after
 integration over all angular variables only two Vandermonde
determinants at the ends of the interval remain:
$\Delta(z(0))\Delta(z(\beta))$ and assure the antisymmetry of wave
functions. In the case of the periodic boundary conditions, one should
be more careful and use  $\delta$-function to match the eigenvalues:
$$
\eqalign{
f(z(\beta)) = \int \prod_k dz_k(0) \Delta^2(z(0))\ f(z(0))
\delta(z(0) - z(\beta))
}
\eqno{(3.9)}
$$
Substituting eqs. (3.1), (3.6) and (3.8) in eq.(2.1) we obtain after
 integration over $A(t)$ a functional integral over eigenvalues
$$
\eqalign{
&{\cal Z}_N(\beta , \lambda ) = {1\over N!}\sum_{\{\cal P\}}
(-1)^{\cal P}
\int \prod ^N_{k=1} {\cal D}z_k(t)
e^{-S_0(z_k)} \cr &\sum_R d_R 
 tr_R\bigg\{ \widehat T {\rm exp} \bigg( {1\over 4N} \int ^\beta _0 dt
 \sum _{i < j}
{(\hat \tau ^R_{ij} \hat \tau ^R_{ji}) \over (z_i -z_j)^2}\bigg)
~{\cal P}\bigg\}
}
\eqno{(3.10)}
$$
where ${1\over N!}\sum_{\{{\cal P}\}}(-1)^{\cal P}$ is a standard
anti-symmetrizator, which appeared because of the skew-symmetry of the
Vandermonde determinant:
$$
\eqalign{
\Delta({\cal P}z(0){\cal P}^{-1})=
(-1)^{\cal P}\Delta(z(0))
}
\eqno{(3.11)}
$$
and $S_0$ is the representation independent part of the action:
$$
\eqalign{
S_0(z) = N\int ^\beta _0 dt \big[ {\dot z^2 \over 2} + V(z)\big]
}
\eqno{(3.12)}
$$
The $\widehat T$-ordering acts in the space of a representation R
$(\tau ^R_{ij}\tau ^R_{ji}$ is the matrix product in this space). As
it follows from eq. (3.2), the action does not contain a quadratic
part for the diagonal elements of the field $A_{ij}$. The integration
over them gives rise to a selection rule for representations.  It is
shown in the Appendix A that it selects only such $R$ for which the
sum of all components of the highest weight,

$$
\eqalignno{
R & = [m_1, m_2 \cdots m_N] \cr
  & m_1 \geq m_2 \cdots \geq m_N,  & (3.13)
}
$$
equals to zero
\footnote\dag{Let us note here that this class of allowed representations is
much wider than the class of the self-conjugate ones (i.e. such that
 $m_{N-n} = N-m_n$) which was considered in the paper [3].}:
$$
\eqalign{
\sum^N_{k=1} m_k = 0
}
\eqno{(3.14)}
$$
     
We conclude from eq. (3.10) that ${\cal Z}_N(\beta ,\lambda )$ can be
represented in the form of  Gibbs partition function:
$$
\eqalign{
{\cal Z}_N(\beta , \lambda ) = \sum_R d_R Tr_R e^{-\beta {\widehat H}_R}
}
\eqno{(3.15)}
$$
where the Hamiltonians $\widehat H_R$ are given by
$$
\eqalignno{
H_R & = P_R \sum^N_{k=1} \big[ -{1 \over 2N} {\partial ^2 \over
\partial z_k^2} + NV(z_k) \big] + {1 \over 4N} \sum_{i < j} {{\widehat
\tau }^R_{ij} {\widehat \tau }^R_{ij} \over (z_i - z_j)^2} & (3.16) }
$$
$P_R$ defined by eq. (A5) is a projector onto the subspace of all
zero weight vectors in the space of  representation R ({\sl i.e.} the
kernel space of the generators of the Cartan subalgebra).  The sum in
eq. (3.15) runs over all irreducible representations obeying the
selection rule (3.14). Only for such representations this subspace is
not empty.
     
We see that unlike the simplest case of the trivial representation the
 eigenvalues cannot be considered here as non-interacting fermions
 (see Appendix A where transformational properties of $\Psi$-functions
 under permutations of eigenvalues are described).
     
A natural question arises whether, for higher representations, the
same double scaling limit is applicable as for the trivial one, since
the interaction among eigenvalues could completely change the critical
properties of the system.  However, we hope that, if the higher
representations really correspond to the implementation of vortices,
the critical properties of the world sheet of string are still the
same as without vortices, at least far away from a vortex position.
So, it is conceivable to adopt the same double scaling procedure as in
[1].  It is based on the fact that the potential V(z) has a shape as
shown in fig.1.  It has a well filled by eigenvalues, which are kept
there since the tunneling through the barrier is exponentially small
with respect to its width.  The decay of this system is not essential
at least in every order of the ${1 \over N}$ (topological) expansion.
The instability corresponding to the dominance of large graphs
(thermodynamical limit) emerges when the eigenvalues near the Fermi
level begin to spill over the top of the barrier, which can be achieved
by tuning either $\lambda $ or N.

\bigskip
    \vskip 25pt

\hskip 70pt
\epsfbox{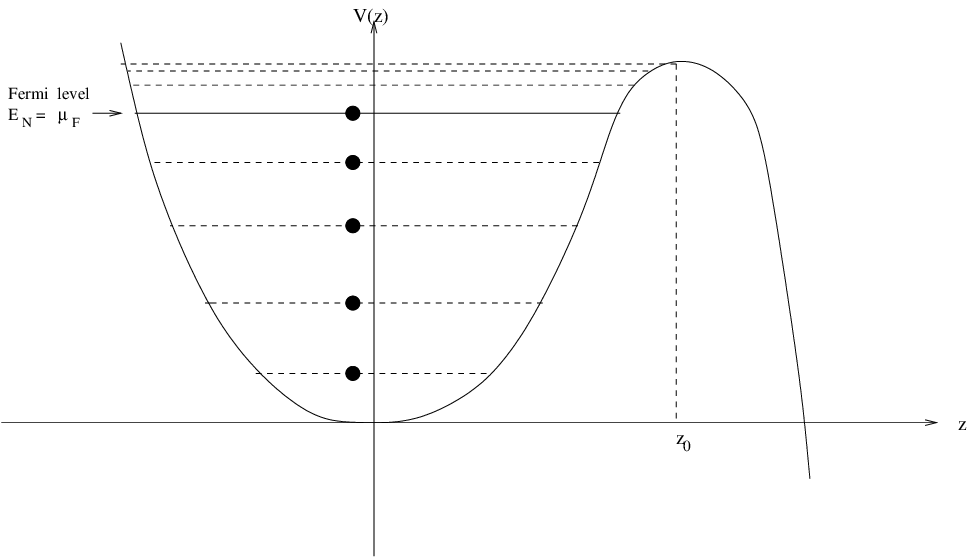}
\vskip 5pt
{\it Fig. 1:  A typical potential $V(z)$ for an eigenvalue $z$ in 
eq.(3.12).  The ground state is fermionic,  i.e.  first $N$
quasi-stable levels in this unstable potential are occupied. 
The criticality appears when the Fermi level $\mu_F$ touches the 
local maximum at $z=z_0$.  }

    \bigskip

This shows that one has to concentrate oneself on the vicinity of the
top which defines all the interesting physics.  Let us choose  new
coordinates
$$
\eqalign{
x_i = \sqrt{N} (z_i - z_0)
}
\eqno{(3.17)}
$$
where $z_0$ is the position of the top ($z_0 = {1 \over \lambda }$
for the cubic potential (2.2)).  Then
$$
\eqalign{
V(z(x)) =
-{1 \over 2N} x^2 + \sum_{k \geq 3} {C_k \over N^{k/2}} x^k
}
\eqno{(3.18)}
$$
all the powers of $X^k, k \ge 3$, are suppressed by the negative
powers of N and can be omitted in the double scaling limit.  Therefore,
the essential part of the Hamiltonian (3.16) is
$$
\eqalign{
H_R = P_R \sum^N_{k = 1} \big[ -{1 \over 2}
{\partial ^2 \over \partial x_k^2} - {1 \over 2} x^2_k \big]
+ {1 \over 2} \sum_{i \neq j}
{{\widehat \tau }^R_{ij}{\widehat \tau }^R_{ij} \over (x_i - x_j)^2}
}
\eqno{(3.19)}
$$
     
Of course, this Hamiltonian is ill defined because of the unstable
quadratic potential and at some moment we have to remember about
higher powers of $x$ in eq. (3.18).  The cubic term in eq. (3.18)
provides an infinite wall at the distance
$$
\eqalign{
\Lambda  \sim \sqrt{N}
}
\eqno{(3.20)}
$$
from the top of the potential.
     
As was noticed in ref. [1], one can define a completely stable system
having the same 1/N-expansion by putting two symmetric walls at
distances $\pm \Lambda $ from the top, {\sl i.e.}  our system can be
stabilized by putting it into the N-dimensional hyper-cubic box $\{
-\Lambda < x_i < \Lambda ,\ i=1, \cdots N\}$.  The convenience of the
eigenvalue representation (3.19) is based on the fact that every
eigenvalue has independent boundary condition.  However, unlike the
singlet representation, the eigenvalues are now interacting.
     
On the other hand, it is obvious that eq. (3.19) can be viewed as the
eigenvalue representation of the matrix upside-down oscillator with
the Hamiltonian:
$$
\eqalign{
{\widehat H}(\varphi) = -{1 \over 2} tr
\big( {\partial ^2 \over \partial \varphi ^2} + \omega ^2 \varphi
 ^2\big)
}
\eqno{(3.21)}
$$
where we have introduced the frequency $\omega $ by simple rescaling
of the time.  On the first glance, eq. (3.21) represents the system
 of $N^2$ non-interacting oscillators
$$
\eqalign{
H(\varphi ) = -{1 \over 2} \sum_{i>j}
 \big( {\partial ^2 \over \partial \varphi _{ij} \partial \varphi
^*_{ij}} \big)
-{1 \over 2} \sum^N_{k=1}
\big({\partial ^2 \over \partial \varphi ^2_{kk} } \big)
}
\eqno{(3.22)}
$$
     
But the boundary conditions, which are determined now by higher terms
of the type
$$
\eqalign{
tr\varphi ^3 = \sum_{i,j,k} \varphi _{ij} \varphi _{jk}
\varphi _{kj}
}
\eqno{(3.23)}
$$
mix all matrix elements and are not so simple as those for
eigenvalues.
     
\chapter{4. Oscillatorial partition functions in different
representations.}

The results of the previous section show that we can view one
dimensional bosonic string theory as a theory of the upside down
matrix oscillator, eq. (3.21), with U(N)-symmetric stabilizing walls
(cut-off) at a large distance $(\sim \Lambda )$ from the top of the
potential.
     
Formally, if we forget for a while about the cut-off, this system is
related to the standard oscillator with the stable matrix potential by
simple analytical continuation to the imaginary frequency:
$$
\eqalign{
\omega  \rightarrow i\omega
}
\eqno{(4.1)}
$$
     
We might hope that physical quantities for the standard oscillator
such as the partition function, Green's functions etc.  already
contain some information about the upside down oscillator.
     
But if we take, for example, the partition function for the standard
oscillator, which is equal to the product of partition functions of
all matrix elements:
$$
\eqalign{
Z^{(N)}(q) = \bigg( {q^{1/2} \over 1-q}\bigg)^{N^2}
}
\eqno{(4.2)}
$$
where
$$
\eqalign{
q = e^{-\beta \omega }\ \ ,
}
\eqno{(4.3)}
$$
we see that it is too simple to be able to describe all the complexity
of physics of 1-dimensional strings even after an analytical
continuation.  Apparently, the information about the cut-off is lost
in (4.2), and we have to try more suitable quantities in order to get
some information about the upside-down oscillator.
     
A natural idea arises to classify all the states of the oscillator
with respect to the irreducible representations of the U(N) rotations
of the matrix coordinate:
$$
\eqalign{
\varphi  \rightarrow \Omega ^+ \varphi \Omega
}
\eqno{(4.4)}
$$
The corresponding wave functions $\Psi _R(\varphi )$ of these states
should transform as matrix elements in a given representation R.
Being formulated as a function of eigenvalues of $\varphi $ this wave
function obeys the Schr\"odinger equation:
$$
\eqalign{
\widehat H_R \Psi _R (z) = E \Psi _R(z)
}
\eqno{(4.5)}
$$
with $\widehat H _R$ given by eq.(3.19).
     
The partition function in the R'th representation will be given by
$$
\eqalign{
Z^{(N)}_R = Tr e^{-\beta \widehat H_R}
}
\eqno{(4.6)}
$$
     
As we shall see in the next section, this quantity contains much more
information about the continuation to the upside-down oscillator.  The
success of this approach is based on the fact that the cut-off wall
for the upside-down oscillator is defined by higher U(N) invariant
terms in the matrix potential (such as $\lambda _1tr \varphi ^3 +
\lambda _2 tr \varphi ^4 + \cdots )$, and therefore it does not
destroy the classification of states in representations: every wave in
a given representation R being sent to the wall will be reflected with
the same properties under U(N) transformations.  Hence, we think that
the analytical continuation procedure should be separate for every
representation.
     
Let us now work out the framework of finding $Z^{(N)}_R$
for the standard oscillator.
     
The most elegant way to do it is to define the so called
twisted partition function:
$$
\eqalignno{
Z^{(N)}(\Omega ) & = Tr (e^{-\beta \widehat H} \widehat \Theta (\Omega
 ) ) = \int de\rho _\Omega (E) e^{-\beta E} & (4.7) }
$$
where $\widehat \Theta(\Omega )$ is a twist operator which ``rotates''
the final states with respect to the initial ones by the U(N) matrix
$\Omega $ and
$$
\eqalign{
\rho _\Omega (E) = {1\over \pi } Im \sum_{E'} \int d^{N^2}\varphi
\
{\Psi ^*_{E'} (\varphi )\Psi_{E'} (\Omega ^+\varphi \Omega ) \over
 E-E' + i\varepsilon}
}
\eqno{(4.8)}
$$
can be called the ``twisted density of states'' though it does not
have even  normal properties of a density (say, it is not
positively defined).
     
Then we can represent the eq. (4.7) as
$$
\eqalign{
Z^{(N)}(\Omega ) = \sum_R d_R Z^{(N)}_R\hchi _R(\Omega )
 }
\eqno{(4.9)}
$$
and for the twisted density of states:
$$
\eqalign{
\rho _\Omega (E) = \sum _Rd_R \rho _R(E) \hchi _R(\Omega  )
}
\eqno{(4.10)}
$$
where the characters $\hchi _R(\Omega )$ obey the orthogonality relation:
$$
\eqalign{
\int (d\Omega ) \hchi _{R_1}(\Omega ^+)\hchi _{R_2}(\Omega  \cdot
 \omega ) =
\delta _{R_1R_2}\hchi _{R_1}(\omega )
}
\eqno{(4.11)}
$$
and hence
$$
\eqalign{
Z^{(N)}_R = \int (d\Omega ) \hchi _R(\Omega )Z^{(N)} (\Omega )
}
\eqno{(4.12)}
$$
and
\def\hrho{\raise3pt\hbox{$\rho$}}
$$
\eqalign{
\hrho _R^{(N)}(E) = \int (d\Omega ) \hchi _R(\Omega )\hrho _\Omega (E)
}
\eqno{(4.13)}
$$
     
The simplest way to make these formulae more explicit is to use
the Green function  defined by the equation:
$$
\eqalign{
\big[ {\partial \over \partial \beta } - tr ({1 \over 2}
{\partial ^2 \over \partial \varphi ^2} - {\omega ^2 \over 2}
\varphi ^2)\big]
G(\beta ,\varphi ,\varphi ') = 0
}
\eqno{(4.14)}
$$
with the initial condition:
$$
\eqalign{
G(0, \varphi , \varphi ') = \delta ^{(N^2)}(\varphi  - \varphi ')
}
\eqno{(4.15)}
$$
     
The solution is well known to be
$$
\eqalignno{
&G(\beta , \varphi ,\varphi ') =
\big( {\omega  \over 2\pi  sh \omega \beta }\big) ^{N^2/2}
 {\rm exp}\big\{ -{\omega  \over 2}{\rm cth} \omega \beta\
tr(\varphi ^2 + \varphi '^2) +
 {\omega  \over sh \omega \beta } tr (\varphi \varphi ')\big\}
&(4.16)
}
$$
     
After the simultaneous change
$$
\eqalign{
\omega  \rightarrow i\omega
}
\eqno{(4.17)}
$$
$$
\eqalign{
\beta  \rightarrow i\beta
}
\eqno{(4.18)}
$$
eq.(4.16) can be viewed as the Green's function for the upside-down
matrix oscillator.
     
Eq.(4.16) appears to be a very useful starting point to many problems
related to the one-dimensional strings.  Let us, for example,
demonstrate that the singlet sector of this theory, {\sl i.e.} the
sector bound to operators of the type
$$
\eqalign{
O_\alpha (t) = tr e^{\alpha \varphi (t)} = \sum^N_{k=1} e^{\alpha
 z_k(t)}
}
\eqno{(4.19)}
$$
which are invariant functions of $\varphi $ depending only on
eigenvalues, can be described in terms of non-interacting fermions.
     
Indeed, if we calculate the two-point function, $D_{\alpha _1\alpha
 _2}(\beta )$,
 we obtain:
$$
\eqalignno{
& D_{\alpha _1\alpha _2}(\beta ) = \ll tr e^{\alpha _1\varphi (0)}tr
 e^{\alpha _2\varphi
(\beta )}\gg = \cr
& \int d^{N^2}\varphi \int d^{N^2}\varphi ' tr e^{\alpha _1\varphi } tr
e^{\alpha _2\varphi '} G(\beta , \varphi , \varphi ') = \cr
& \big( {\omega  \over 2\pi sh \omega \beta }\big) ^{N^2 \over 2}
 \sum_{k,m}
\int \prod ^N_{i=1} dz_idz_i' {\rm exp}
\left\{-{\omega  \over 2}cth \omega \beta (z^2_i + z_i^{'2}) \right\}
 \cr
& \times e^{\alpha _1z_k + \alpha _2z'_m} \times \Delta ^2(z)\Delta
 ^2(z')
\int (d\Omega ){\rm exp}\big\{ {\omega  \over sh \omega \beta }
 \sum_{\alpha ,b}
\Omega ^+_{ab}z'_b\Omega _{ba}z_a\big\} \cr
& = \big( {\omega  \over 2\pi sh \omega \beta }\big) ^{N^2 \over 2}
\sum_{k,m} \int \prod ^N_{i=1}dz_idz'_i {\rm exp}
\big\{ -{\omega  \over 2}cth \omega \beta (z^2_i + z'^{2}_i) +
{\omega  \over sh\omega \beta }z_iz_i' \big\} \cr
& \times \Delta (z) \Delta (z') e^{\alpha _1z_k + \alpha _2 z'_m} &
(4.20)
}
$$
where we used the ``angular'' decomposition of the matrices $\varphi =
\omega ^+z\omega $ and \hfill \break $\varphi '=(\Omega \omega
)^+z'(\Omega\omega )$ and the well-known formula [16] :
$$
\eqalignno{
\int (d\Omega ) & {\rm exp} a{\rm tr}(\Omega ^+z'\Omega z) =
\big(\prod ^{N-1}_{k=1} k!\big) a^{-{N(N-1) \over 2}}
{{\rm det }_{km}(e^{a z'_k z_m}) \over \Delta (z') \Delta (z)} &(4.21)}
$$

We see from eq. (4.20) that the initial and final states in terms of
eigenvalues appear to be completely anti-symmetric due to the
Vandermonde determinants $\Delta (z')$ and $\Delta (z)$. Hence, they
can be represented as Slater determinants of the eigenfunctions of N
oscillators (or upside-down oscillators):
$$
\eqalign{
\Psi _{n_1n_2 \cdots n_N}(z_1, \cdots z_N) =
{1 \over \sqrt{N!}} \underrel{det}{(i,j)} \Psi _{n_i}(z_j)
}
\eqno{(4.22)}
$$
     
So, we deal with free fermions.  In principle, starting from here we
can obtain all the results of the paper [11] for the two-point function
in the double scaling regime.
     
On the other hand, if we take the two point function of another type
$$
\eqalignno{
K^{(\beta )}_{\alpha _1\alpha _2} &= < {tr \over N} \big[ O_{\alpha
 _1}(0)
O_{\alpha _2}(\beta )\big] > \cr &=
\int d^{N^2}\varphi d^{N^2}\varphi ' {tr \over N} \big[ e^{\alpha
 _1\varphi}
e^{\alpha _2\varphi '}\big] G(\beta ,\varphi ,\varphi ')
&{(4.23)}
}
$$
we  have a more complicated integral over U(N)
group of the type
$$
\eqalign{
\int (d\Omega ) \Omega ^+_{ij} \Omega _{ji} {\rm exp}
\big\{a \ tr(\Omega ^+z'\Omega z)\big\}
}
\eqno{(4.24)}
$$
     
Since the matrix element of the adjoint representation
$D^{k\ell }_{ij}$ looks as
$$
\eqalign{
D_{k\ell }^{ij} = (\Omega ^+)_{ji} \Omega _{k\ell }
- {1\over N} \delta_k^i \delta^j_\ell
}
\eqno{(4.25)}
$$
we expect that the Green's function (4.23) describes the propagation
of states transforming under the adjoint representation of U(N).
Indeed, one can show (see Appendix B) that
$$
\eqalign{
K_{\alpha _1\alpha _2}(\beta ) = \sum_{m, n} \langle 0 \vert(e^{\alpha _1z_m})
(e^{-\beta {\widehat H}_{adj}})_{mn}^{mn}
 \left( e^{\alpha _2z_n}\right) \vert 0\rangle}
\eqno{(4.26)}
$$
where
$$
\eqalignno{
&({\widehat H}_{adj})^{mn}_{ij}  = \delta^{mn} \delta_{ij}\sum_k
\big( -{1 \over 2}{\partial ^2 \over \partial z^2_k} -
{\omega ^2 \over 2}z^2_k \big) + \cr
& + {1\over 2}\delta^{mn} \delta_{ij}
\bigg(\sum_{k\ne i}{1\over (z_i - z_k)^2} +
 \sum_{k\ne n}{1\over (z_k - z_n)^2} \bigg) -
{\delta^m_i \delta_j^n \over (z_i - z_j)^2}
& (4.27)
}
$$
which coincides with eq. (3.16) if we take $\widehat \tau ^{adj}_{ij}$
there.  The simplest of these quantities is
$$
\eqalign{
K_0(\beta ) = {\partial ^2 \over \partial \alpha _1\partial \alpha _2}
K_{\alpha _1\alpha _2}(\beta ) \vert _{\alpha _1 = \alpha _2 = 0}
= <tr [\varphi (0)\varphi (\beta )] >
}
\eqno{(4.28)}
$$

\bigskip
    \vskip 5pt
    
    %
\hskip 70pt
\epsfbox{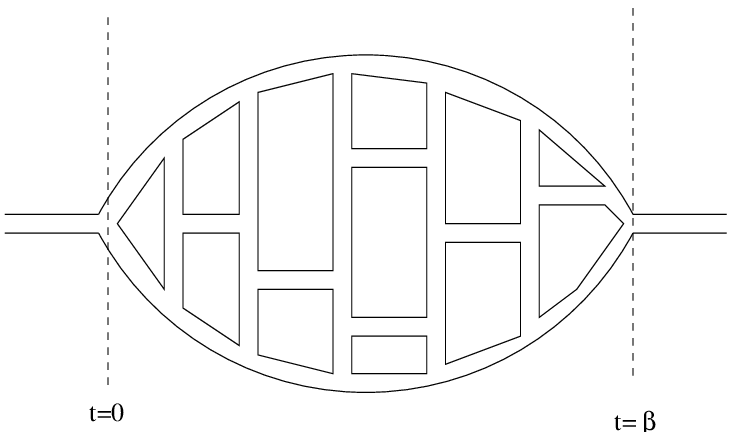}
\vskip 5pt
{\it  Fig. 2: A typical $\varphi^3$ Feynman graph for the adjoint 
propagator (4.28). It has a topology of a disc with two points on
the boundary fixed. }
 
    \bigskip

In the case of the matrix $\varphi ^3$-potential, $K_0(\beta )$ will
 be defined by the diagrams of the type drawn in fig.2.  One can see
 that it represents an open string amplitude with  free boundary
 conditions and two points on the boundary fixed in the t-space at 0
 and $\beta $.

On the other hand, this diagrammatic picture allows us to interpret
the partition function (4.12) in the adjoint representation directly
as describing the one-vortex-anti-vortex-pair sector of the model.
     
Indeed, open string is the only intermediate state propagating in the
adjoint representation.  For $Z_{adj}$ we have the diagrammatic
representation drawn in fig.3.  It looks similar to that of the
correlator (4.28)) (fig.2) but, instead of having ``in'' and ``out''
states at the ends, corresponding to some particular operators
$\varphi (0)$ and $\varphi (\beta )$, it is continued periodically
looking now as being wrapped around a cylinder.  In fact, there are
two loops in the graph which wrap around the time circle of the length
$\beta $, in  opposite directions with respect to each other (if we
introduce, say, the clock-wise orientation for every loop).  As we
mentioned in section 2 it is exactly the definition of vortex and
anti-vortex encircled by these two loops, correspondingly.  All other
loops are small in the $t$-space and do not contain vortices.

\bigskip
    \vskip -20pt
    
    %
\hskip -180pt
\epsfbox{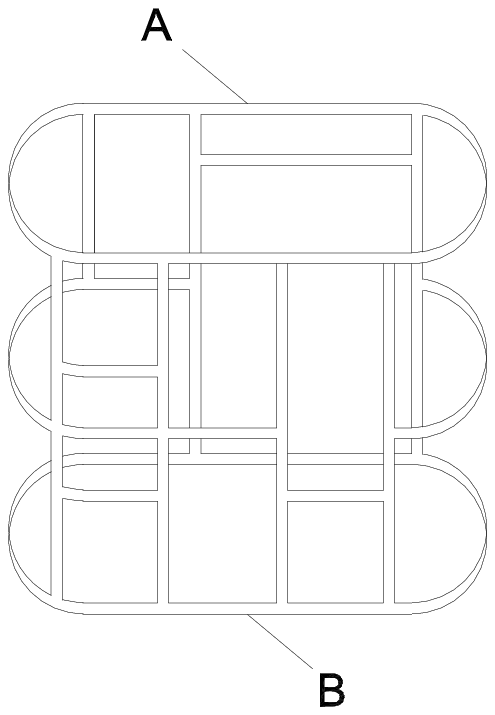}
\vskip -20pt
{\it Fig. 3: A typical $\varphi^3$ Feynman graph for the adjoint 
partition function $Z_{\rm adj}$. It has the topology of a cylinder,
which is wrapped around the circle of the radius $\beta/2\pi$. }
 
    \bigskip

It is possible to generalize this picture to higher representations.
     
From these considerations we draw the conclusion that higher
representations have to be related in some way to open strings with
the free (non-renormalized by some additional particles at the ends)
boundary conditions and, hence, to bigger numbers of vortices.
     
Now let us proceed with the partition functions.  The untwisted
partition function can be defined as:
$$
\eqalign{
Z^{(N)}(\beta ) = \int d^{N^2}\varphi G(\beta , \varphi , \varphi )
}
\eqno{(4.29)}
$$
Using eqs.(4.16) and (4.29) we can easily reproduce the trivial
result (4.2).
     
The twisted partition function is equal to
$$
\eqalign{
Z^{(N)}(\beta ,\Omega ) = \int d^{N^2}\varphi G(\beta ,\varphi ,\Omega
 ^+\varphi \Omega )
}
\eqno{(4.30)}
$$
Here we simply have rotated the final states with respect to the
initial ones by a matrix $\Omega $.
     
Let us first consider the trivial representation (singlet).  Using
eqs. (4.12) and (4.16) and diagonalizing again the matrix $\varphi $,
we integrate over the angular variables by eq.(4.21) and obtain:
$$
\eqalignno{
 Z_{sing}^{(N)} (\beta ) 
& = \big( {\omega  \over 2\pi sh\omega \beta }\big)^{N/2} \int \prod
 ^N_{k=1}
dz_k{\rm exp}\big\{-{\omega  \over 2}cth \omega \beta z^2_k\big\} \times
 \cr
& \quad \quad \times {{\rm det}_{ij}}
\big[ {\rm exp}\big\{{\omega  \over sh\omega \beta }z_iz_j\big\} ]
& (4.31)
}
$$
It is obvious from here that $det$ in eq. (4.31) produces the expected
fermionic statistics for $\{z_k\}$.  Say, the contribution of
configurations with coinciding coordinates $z_m = z_n$ for some $m$
and $n$, is zero.  Later we shall compute $Z_{sing}$ by other means
and reproduce well known results for non-interacting fermions.  The
measure in eq.(4.30) is invariant under the group transformations
$\varphi \rightarrow \omega ^+\varphi \omega $.  We can use this
invariance to show that $Z^{(N)}(\beta ,\Omega )$ depends only on the
eigenvalues of $\Omega $.  Indeed, we can use the $\omega
$-transformation of $\varphi $ to diagonalize $\Omega $, since nothing
depends on $\omega $:
$$
\eqalign{
(\omega \Omega \omega ^+)_{kj} = \delta _{kj}e^{i\theta _j}
}
\eqno{(4.32)}
$$
     
Let us perform now the gaussian integration over $\varphi $
in eq.(4.30).
  We obtain the useful formula:
\footnote\dag{We thank M. Douglas who showed us this formula.  
He claimed it was known long time ago but we did not find any
references.}
     
$$
\eqalign{
Z^{(N)} (\beta , \theta ) = q^{{N^2 \over 2}} \prod ^N _{k, m=1}
{1 \over 1 - qe^{i(\theta _k - \theta _m)}}
}
\eqno{(4.33)}
$$
From eqs. (4.12), (4.30) and (4.33) we obtain:
$$
\eqalign{
Z_R (\beta ) = {1 \over N!} \int ^{2\pi }_0  \prod ^N _{k=1 }
{d\theta_k \over 2\pi } \mid \Delta (e^{i\theta }) \mid ^2 \hchi _R(e^{i\theta
 })
\prod ^N _{k, m=1}
\big[ {e^{-{\omega \beta  \over 2}} \over 1 - {\rm exp}[-\omega \beta
 +i(\theta _k-\theta _m)]}\big]
}
\eqno{(4.34)}
$$
where
$$
\eqalign{
\Delta (e^{i\theta }) = \prod _{k>m} (e^{i\theta _k} - e^{i\theta _m})
}
\eqno{(4.35)}
$$
and the character $\hchi _R$ is given by the Weyl formula:
$$
\eqalign{
\hchi _R(e^{i\theta}) = {\det _{km} (e^{il_k\theta _m}) \over \Delta
 (e^{i\theta })}
}
\eqno{(4.36)}
$$
     
Using the complex variables
$$
\eqalign{
\xi _k = e^{i\theta _k}
}
\eqno{(4.37)}
$$
eq. (4.34) can be written in the form of the contour integrals:
$$
\eqalignno{
Z_R(q) & = {q^{N^2 /2} \over N!} \ointop \cdots
\ointop \prod ^N_{k=1} {d\xi _k \over \xi _k} \mid \Delta (\xi ) \mid
 ^2
\hchi _R(\xi ) \times \cr
& \times \prod ^N_{m,k=1} (1-q\xi _ m{\bar \xi }_k) ^{-1}
& (4.38)
}
$$
The integrals are taken over the unit circle, and all the poles inside
 this circle are concentrated either at the origin or on the circle of
 the radius $q$.
     
In principle, we can obtain by this formula using the direct
computation some results for given R and N.  However, we need the
general formula for arbitrary N.
     
 Let us work out a more effective method for it.
     
Using the "generating function for characters" [17]:
$$
\eqalign{
\prod ^N_{k,l=1} {1 \over 1-\xi _k\eta _l} = \sum_R \hchi _R(\xi )\hchi
 _R(\eta )
}
\eqno{(4.39)}
$$
and the fact that the character is a homogeneous function:
$$
\eqalign{
\hchi _R(p\xi ) = p^{\sum^N_{k=1}m_k} \hchi _R(\xi )
}
\eqno{(4.40)}
$$
where $\{ m \}$ are again the highest weight components, we can
represent eq. (4.34) as:
$$
\eqalign{
Z_R^{(N)}(q) = \sum_{\{R'=\ell _1 \cdots \ell _N\}} M^{R'}_{R \otimes R'}
\quad q^{\sum^N_{k=1}(\ell_k + {1 \over 2})}
}
\eqno{(4.41)}
$$
where the integer numbers $\{ \ell _k \}$ obeying the condition
$$
\eqalign{
\ell _1 > \ell _2 \cdots > \ell _N
}
\eqno{(4.42)}
$$
are related to the components $\{ n_k\} $ of the highest weight of
$R'$ by the formula
$$
\eqalign{
\ell _k = n_k + N - k
}
\eqno{(4.43)}
$$
and
$$
\eqalign{
M^{R''}_{R \otimes R'} = {1 \over N!} \int ^{2\pi } _0 \prod ^N_{k=1}
{d\theta _k \over 2\pi } \vert \Delta (e^{i\theta}) \vert^2 
\hchi _R(e^{i\theta }
 )\hchi_{R'} (e^{i\theta})\overline{\hchi}_{R''}(e^{i\theta })
}
\eqno{(4.44)}
$$
is the multiplicity of a representation $R''$ in the tensor product $R
\otimes R'$ of irreducible representations $R$ and $R'$.  For our
purposes we need only the case $R' = R''$. It is important to note
that only non-negative values of $\ell_k$ appear in eqs. (4.41), which
makes the sum convergent.  For the trivial (singlet) representation,
we have $\hchi _{\rm sing}(e^{i\theta})=1$, and eqs.(4.41)-(4.44)
give:
$$
\eqalignno{
Z^{(N)}_{\rm sing}(q) & = \sum^\infty _{\ell _N=0} \sum^\infty _{\ell
 _{N-1}=
\ell_N+1} \cdots \sum^\infty _{\ell _1 = \ell _2 +1}
q^{\sum^N_{k=1}(\ell _k + {1 \over 2})} = \cr
& = {q^{N^2/2} \over (1-q)(1-q^2) \ldots (1-q^N)}
& (4.45)
}
$$
     
We have obtained the partition function of N fermions in the harmonic
potential, as it was expected.  The fermionic statistics appears here
because of the ordering condition (4.42) for the components of the
highest weight (see ref. [18] for some analogies).
     
The formula (4.41) gives another check of the selection rule mentioned
in the \- section 3: $Z_R$ is nonzero only for representations $R=\{
m_k\}$ with the sum of highest weight components equal to zero
(eq.(3.14)).
     
If the conditions
$$
\eqalign{
n_k-n_{k+1} \geq \sum^N_{j=1} (m_j + \mid  m_N \mid ) = N\mid m_N \mid
}
\eqno{(4.46)}
$$
(where $\{n_k\}$ are the components of the highest weight of $R'$ in
eq. (4.41)) are fulfilled for all $1 \leq k < N$, then [17]:
$$
\eqalign{
M^{R'}_{R \otimes R'} = d^{(0)}_R
}
\eqno{(4.47)}
$$
where $d^{(0)}_R$ is the dimension of the subspace of all zero weight
vectors in the space of a representation $R$ defined by eq. (A6) and
in general
$$
\eqalign{
M^{R'}_{R \otimes R'} \le d_R^{(0)}
}
\eqno{(4.48)}
$$
Eq. (4.47) means that, for $\beta \rightarrow 0 $ (or $q=e^{-\omega
\beta } \rightarrow 1$ ),
$$
\eqalign{
Z_R(\beta ) \underrel{\sim}{\beta \rightarrow 0} {d^{(0)}_R \over
(\omega \beta )^N} }
\eqno{(4.49)}
$$
     
The general structure of the partition function $Z_R^{(N)}(q)$ is such
that
$$
\eqalign{
Z^{(N)}_R(q) = P^{(N)}_R(q) Z_{\rm sing}^{(N)}(q)
}
\eqno{(4.50)}
$$
where $Z^{(N)}_{\rm sing}(q)$ is the singlet partition function
defined by eq. (4.45) and $P_R^{(N)}(q)$ is a polynomial in $q$ of a
degree not higher than $N \cdot m$, where $m$ is the sum of positive
(or negative) components of the highest weight of $R$.
     
From eq. (4.49) we have:
$$
\eqalign{
P_R^{(N)}(1) = d_R^{(0)}
}
\eqno{(4.51)}
$$
$P_R^{(N)}(q)$ can be regarded as the $q$-analog of the multiplicity
of the zero weight and is a known object in modern group theory
\footnote\dag{We are
indebted to A.Kirillov for drawing our attention to this fact.}.  They
appear to be a particular case of the Kostka-Green-Foulkes polynomials
[19].  We elaborated an effective recurrence algorithm suitable for us
(see Appendix C for details) which enabled us to obtain for two
infinite series of Young tables
$$
A_n = (n,0, \ldots \ldots ,-1,-1, \ldots,-1)
\eqno{(4.52)}
$$
and
$$
B_n = (n,0, \ldots\ldots  ,0,-n)
\eqno{(4.53)}
$$
the following simple answers for the corresponding generating
functions
$$
\eqalign{
{\cal P} ^{(N)}_A(x,q) = \sum^\infty_{n=0} P^{(N)}_{A_n}(q)x^n =
\prod^{N-1}_{k=1} (1+xq^k)
}
\eqno{(4.54)}
$$
$$
\eqalign{
{\cal P }^{(N)}_B(x,q) = \sum^\infty_{n=0} P^{(N)}_{B_n}(q)x^n =
\prod^{N-1}_{k=1} {1 \over 1-xq^k}
}
\eqno{(4.55)}
$$
Eqs. (4.54) and (4.55) have the form of the grand canonical partition
functions for fermions and bosons, respectively, in the $N-1$ level
system where $x$ plays the role of the fugacity. Unfortunately,
answers for other representations cannot be represented in a similar
simple form but, nevertheless, one can obtain for representations
having a small number of non-zero components of their highest weights
formulae valid for arbitrary $N$.

\chapter{5. Partition functions for the upside-down matrix oscillator }

Now let us try to generalize the methods of the previous section to
a more complicated case of the upside down oscillator.
     
The major complification is the necessity to introduce a cut-off in
order to stabilize this unstable system. Any cut-off decreases the
high symmetry of the oscillator and makes an exact solution
impossible. The only tool in our hands is the analytical continuation
in the frequency but we should be ready to confront with divergences
and, hence, we need additional guesses in order to get physically
reasonable answers.
     
In order to classify the partition functions with respect to U(N)
representations, we have to define again the twisted partition
function.  As it was noted in the section 4, we can formally obtain it
changing $\omega \rightarrow i\omega $ in the Green function $G(\beta
,\varphi ,\varphi ')$.  Then, after the Gaussian integration over
$\varphi $ in eq. (4.30), we formally get the twisted partition
function for the upside-down oscillator
$$
\eqalign{
Z^{(N)}(\beta , \theta)=e^{-i\omega \beta {N^2 \over 2}}\prod
 ^N_{k,m=1}
\big[1 - e^{-i\omega \beta +i(\theta_k - \theta_m)}\big]^{-1}
}
\eqno{(5.1)}
$$
     
Instead of trying to give a meaning to eq.(5.1), let us make an
attempt to obtain $Z^{(N)}(\beta , \theta)$ solving directly the wave
equation
$$
\eqalign{
\widehat H \Psi (\varphi ) = \epsilon \Psi (\varphi )
}
\eqno{(5.2)}
$$
for $\widehat H$ taken from eq.(3.22).
     
If we forget for a while about the boundary conditions (i.e. the
cut-off) we can consider eq. (5.2) as a collection of independent wave
equations for off-diagonal matrix elements:
$$
\eqalignno{
-{1 \over 2} \big({\partial ^2 \over \partial \varphi _{ij}\partial \varphi
 ^*_{ij}}
 & + \omega ^2\varphi _{ij}\varphi ^*_{ij}\big)\Psi_{ij}(\varphi _{ij})=\epsilon
 _{ij}\Psi_{ij}
(\varphi _{ij}) & (5.3)\cr
& \quad {\rm for} \quad i<j
}
$$
and diagonal matrix elements:
$$
\eqalignno{
-{1 \over 2} \big( {\partial ^2 \over \partial \varphi ^2_{ii}} & + \omega
 ^2
\varphi ^2_{ii}\big)
\Psi _{ii}(\varphi _{ii}) = \epsilon _{ii}\Psi _{ii}(\varphi _{ii})  &
(5.4)\cr
& \quad {\rm for} \quad i=1,2 \cdots N
}
$$
     
If we introduce the parameterization (dropping for a while the indices
(i,j)):
$$
\eqalign{
\varphi  = \sqrt{r} e^{i\theta}
}
\eqno{(5.5)}
$$
we get the following form of eq. (5.3):
$$
\eqalign{
\hchi ^{''}_m + {2 \over r}\hchi '_m + \big[ \big({\omega  \over 2}\big)^2 +
{\epsilon  \over 2r} - {m^2 -1 \over 4r^2}\big] \hchi  = 0
}
\eqno{(5.6)}
$$
where $\hchi $ is related to $\Psi$ as
$$
\eqalign{
\Psi(r, \theta) = \sqrt{r}\sum^\infty _{m=-\infty } \hchi _m(r)e^{im\theta}
}
\eqno{(5.7)}
$$
Eq.(5.6) coincides with the Schr\"odinger equation for the hydrogen
atom with the energy $\big({\omega \over 2}\big)^2$, the charge
$\epsilon /2$ and the angular momentum $\ell = {\mid m \mid - 1 \over
2}$. The imaginary values of $\omega $ in eqs. (5.3) and (5.4)
correspond to the discrete spectrum of the hydrogen (negative
energies), and we get:
$$
\eqalign{
\epsilon  = \mid \omega  \mid (\mid m \mid + 1 + 2k), \quad
k = 0,1,2 \cdots
}
\eqno{(5.8)}
$$
The case of real $\omega $ in eq.(5.6) (the upside-down oscillator)
corresponds to the continuous spectrum of the hydrogen, and we have to
introduce some cut-off to define it.
     
Again, as in the case of the infinite time interval [6], we have
strong reasons to believe that, for the exact definition of the
spectrum, only the quasi-classical asymptotic of the wave function
suffices:
$$
\eqalignno{
\hchi_m (r) \sim {2\over r} \sin \bigg(r +
{\epsilon  \over 2\omega } log(2r) &- {\pi  \over 2} {\mid m \mid -1
 \over 2}
 + arg \Gamma ({\mid m \mid + 1 \over 2} + {i\epsilon \over 2\omega
 })\bigg)
&(5.9)
\cr
& {\rm for} \quad r \gg {1 \over 2 \omega }(m^2 + {\epsilon ^2 \over
 \omega ^2})
}
$$
     
The Schr\"odinger equation (5.4) for the diagonal elements describes
for an imaginary $\omega $ a collection of standard oscillators having
the equidistant spectrum:
$$
\eqalign{
\epsilon _n = \mid \omega  \mid (n + {1 \over 2}), \quad n=0,1,2 \cdots
}
\eqno{(5.10)}
$$
and for a real $\omega $ the solution of (5.4) is a parabolic cylinder
 function
having the quasi-classical asymptotics
$$
\eqalignno{
\Psi (x) \sim x^{-{1 \over 2}} & {\rm exp}\big\{ \pm i \big[x^2 +
{\epsilon  \over \omega }log x + \cr
& + arg \Gamma (i{\epsilon  \over \omega } + {1 \over 2}) \big] \big\}
& (5.12)
}
$$
valid for $x \gg {\epsilon ^2 \over \omega ^3}$.
     
The problem now is that we are, in principle, not allowed to define
the energy spectrum for every matrix element, $\varphi
_{ij}=\sqrt{r_{ij}}{\rm exp}i\theta_{ij}$ and $\varphi _{ii} = x$,
separately, since we have the U(N) invariant boundary condition, say:
$$
\eqalign{
\Psi (\varphi ) = 0 \quad {\rm for} \quad tr\varphi ^2 \geq \Lambda ^2 \sim N
}
\eqno{(5.13)}
$$
which mixes all matrix elements.
     
Strictly speaking, to solve this problem we have to consider first the
whole wave function for all matrix elements, in the form
$$
\eqalign{
\Psi (\varphi ) = \sum_{\{\epsilon \} \{m \} }C(\{ \epsilon \}) \prod
 _{i>j}
\Psi _{ij}(\varphi _{ij}, \epsilon _{ij})\prod _k\Psi _{kk}(\varphi _{kk},\epsilon
 _{kk})
}
\eqno{(5.14)}
$$
and than find the coefficients $C(\{ \epsilon \})$ from the condition
(5.13).  But we are not able to find them directly.
     
Instead of doing this we shall assume as in ref. [1] that the
essential part of the spectrum is defined by the phase of reflection
from the quadratic potential (last terms in the phases of the
asymptotics (5.9) and (5.12) ) and not by the infinite wall, appearing
in the conditions (5.13).
     
To justify it let us note that every term in the expansion (5.11) has
 the quasi-classical asymptotics of the form
$$
\eqalignno{
& {\rm exp} \pm i \big[ tr\varphi ^2 + { 1\over \omega }\sum_{i \geq j}\epsilon
 _{ij}
\log \varphi _{ij} + \cr
& + \sum_{i>j} arg \Gamma \big({\mid m_{ij} \mid + 1 \over 2} + i
 {\epsilon _{ij} \over 2\omega } \big) +
\sum_k arg \Gamma \big(i{\epsilon _{kk} \over \omega } + {1 \over
 2}\big) \big]
& (5.15)
}
$$
     
The first overall term in the phase of eq. (5.15), $e^{\pm i{\rm
tr}\varphi ^2}$ does not have any influence on the density of states,
since it is invariant under U(N) transformations, $\varphi \rightarrow
\Omega ^+\varphi \Omega $, as well as the boundary condition (5.13).
     
The second logarithmic term, as we hope, will give only slow
dependence on the cut-off $\Lambda $ in the density of states, as it
was in the simpler situation for the eigenvalues of the matrix
$\varphi $ in the singlet representation [4-7].
     
In other words, we expect that the density of spectrum for every
matrix element can be defined from the phase quantization condition.
Hence, from (5.9) we have for the off-diagonal degrees of freedom:
$$
\eqalignno{
{\Lambda^2 \over 2}+
{\epsilon  \over 2\omega } log \Lambda ^2 - {\pi  \over 2} {\mid m \mid
 -1 \over 2}
 + & arg \Gamma \big({\mid m \mid + 1 \over 2} + {i\epsilon  \over
 2\omega }\big)
= \pi n \cr
& n = 0,1,2 \cdots
& (5.16)
}
$$
which gives for the density of states $\rho_o (\epsilon )=
{\partial n \over \partial \epsilon }$
$$
\eqalign{
\rho ^{(m)}_o(\epsilon ) = -{1 \over 2\pi \omega } Re \psi \big(
 i{\epsilon  \over 2\omega } +
{\mid m \mid + 1 \over 2}  \big)
+ {1 \over 2\pi \omega } log \Lambda ^2
}
\eqno{(5.17)}
$$
where
$$
\eqalign{
\psi (x) = -\sum ^\infty _{k=0} \bigg({1 \over x+k}
- {1\over k+1}\bigg) - C
}
\eqno{(5.18)}
$$
is the $\psi $-function of Euler; $C$ is the Euler constant.
     
For the diagonal degrees of freedom we obtain:
$$
\eqalignno{
\Lambda^2 +
{\epsilon  \over \omega } log \Lambda^2  + arg \Gamma \big(i{\epsilon
 \over \omega } + {1 \over 2}\big) & =
\pi n \cr
& n=0,1,2 \cdots
& (5.19)
}
$$
or, for the density of states $\rho _d(\epsilon )=
{\partial n \over \partial \epsilon }$:
$$
\rho _d(\epsilon ) = -{1 \over \pi \omega} Re \psi (i{\epsilon  \over \omega }
 + {1 \over 2}) +
{1\over\pi\omega}\log \Lambda^2
\eqno(5.20)
$$
     
Let us now calculate the twisted partition function for the
 upside-down matrix oscillator.  We shall start from the ``twisted
 density of states'' for the off-diagonal matrix elements.
     
Since $\varphi _{ij} = \sqrt{r_{ij}}e^{i\theta_{ij}}$ ,
 the
twist $\varphi  \rightarrow \Omega ^+\varphi \Omega $ results in
$$
\eqalign{
\theta_{ij} \rightarrow \theta_{ij} + \theta_i - \theta_j
}
\eqno{(5.21)}
$$
and we conclude from eqs. (5.7) and (5.17), that
$$
\eqalignno{
&\rho (\theta,\epsilon ) = \sum^\infty _{m=-\infty }e^{im\theta}\rho ^{(m)}(\epsilon)
 \cr
& = {1 \over \pi } \sum^\infty _{k=0} \sum^\infty _{m=-\infty }
{e^{im\theta}({\mid m \mid + 1 \over 2} + k ) \over ({\epsilon  \over
 \omega })^2 +
(k+ {\mid m \mid + 1 \over 2})^2} +
\delta (\theta)\log \tilde\Lambda ^2 & (5.22)
}
$$
The cut-offs which we have introduced till now are not U(N) invariant.
Therefore, we shall drop in what follows the last term with the
periodic $\delta $-function having in mind that the logarithmic
divergency will appear on the next stages of the calculations and we
will have to recover a cut-off again.
     
The remaining double sum can be easily calculated (taking
$j=\mid m \mid + 1 + 2k $ as a new variable) and the result is:
$$
\eqalign{
\rho (\theta, \epsilon ) = {\sinh {\epsilon  \over \omega }(\pi  - \theta) \over
\sinh {\epsilon  \over \omega }\pi
\sin \theta}
}
\eqno{(5.23)}
$$
     
It is remarkable that the twisted density of states does not contain a
cut-off parameter for $\theta \neq 2\pi m, m =0, \pm 1, \pm 2 \ldots$.
The twisted partition function of a non-diagonal matrix element can be
calculated taking both parts of the spectrum above and below the top
of the potential, i.e. for positive as well as for negative $\epsilon
$:
$$
\eqalignno{
& Z_0(\theta, \beta ) = \int ^{+\infty} _{-\infty } d\epsilon \ {\rm e}^{-\beta
 \epsilon }\rho (\theta, \epsilon ) = \cr
& = {1 \over \sin \theta} \int ^{+\infty} _{-\infty }  d\epsilon \
{\rm e}^{-\beta \epsilon }
{\sinh {\epsilon  \over \omega }(\pi  - \theta) \over 
\sinh {\epsilon  \over \omega } \pi }
& (5.24)
}
$$
For
$$
\eqalign{
  0 < \theta < 2\pi
}
\eqno{(5.25)}
$$
the integral in (5.24) is convergent.  The calculation gives:
$$
Z_0(\theta, \beta ) = {1/2 \over \cos \beta \omega - \cos \theta }
\eqno{(5.26)}
$$
     
For $\theta = 0$ and $2\pi$  we
understand eq. (5.26) as an analytical continuation of eq. (5.24).
     
If we rewrite eq. (5.26) as
$$
\eqalign{
{1/2 \over \cos \beta \omega - \cos \theta  } = { {\rm e}^{-i\beta \omega }
 \over (1 - {\rm e}^{-i(\beta \omega +\theta )})
(1 - {\rm e}^{-i(\beta \omega -\theta )})}
}
\eqno{(5.27)}
$$
we find that this is quite similar to the factor ${1 \over
1-qe^{i(\theta _k-\theta _m)}}$ in eq.(4.33): the differences is only
in the change $\omega \rightarrow i\omega $.  It means that we do not
need to introduce a cut-off for the off-diagonal matrix elements.  For
the diagonal matrix elements $\varphi _{ii}$ , the partition function
does not depend on twist angles but the problem is that the integral
over energies
$$
\eqalignno{
& Z_d(\beta ) = \int ^\infty _{-\infty } d\epsilon\ e^{-\beta \epsilon }
 \rho _d(\epsilon ) = -
\int ^\infty _{-\infty } d\epsilon\ e^{-\beta \epsilon } {1 \over
 \pi } Re \psi
(i{\epsilon  \over \omega } + {1 \over 2})
& (5.28)
}
$$
is highly divergent and needs a cut-off procedure. However, we can
formally deform the contour of integration from the real axis to the
one shown in fig.4 and expand $\psi $-function in pole terms
(eq.(5.18)).  Then we can calculate the integral in eq. (5.28) term by
term:
$$
\eqalignno{
Z_d(\beta ) & = {1 \over \pi } \sum ^\infty _{h=0}
\int  d\epsilon e^{-\beta \epsilon } {n + {1 \over 2} \over {\epsilon
 ^2 \over \omega ^2} +
(n + {1 \over 2})^2} = \sum ^\infty _{n=0} e^{i\omega \beta (n + {1
 \over 2})}
= \cr
& = {1 \over 2\sin \beta \omega /2} &
(5.29)
}
$$

\bigskip
    \vskip -30pt
    
    %
\hskip -180pt
\epsfbox{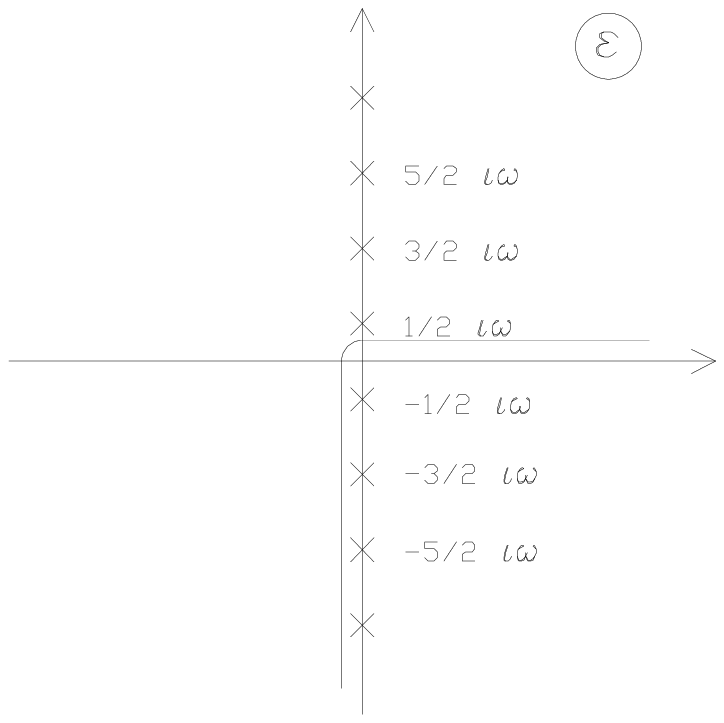}
\vskip -20pt
{\it  Fig. 4:The contour of integration for the diagonal matrix 
elements in the partition function (5.28). }
 
    \bigskip

If we keep assuming that all $N^2$ degrees of freedom $\varphi _{ij}$
for the upside down matrix oscillator are independent , we can write
down the whole twisted partition function as a product of $N^2$
factors (5.26) and (5.29)
$$
\eqalign{
Z^{(N)}(\theta ,\beta ) = \bigg(2\sin{\omega \beta \over 2}\bigg) ^{-N}
 \prod _{i>j}
{1\over \cos \omega \beta - \cos (\theta _i - \theta _j) }
}
\eqno{(5.30)}
$$
which coincides with eq.(5.1), and the partition function in a given
representation R takes the form:
$$
\eqalign{
Z^{(N)}_R(\beta ) = {1 \over N!} \int ^{2\pi }_0 \prod_{k=1}^N
{d\theta
 _k  \over 2\pi }
\mid \Delta  (e^{i\theta }) \mid ^2 \hchi _R (e^{i\theta }) Z^{(N)}(\theta
 ,\beta )
}
\eqno{(5.31)}
$$
It looks exactly the same as eq.(4.34) for the standard oscillator but
with the change $\omega \rightarrow i\omega $.  This is an encouraging
analogy for us if we remember that we used for the derivation of
eq. (5.30) the true spectrum of the upside-down matrix oscillator.
     
Let us also note that in eq. (5.31) there is no factorization to $N^2$
independent degrees of freedom. Effectively, they interact and after
integration over the angles only N effective degrees of freedom remain
({\sl e.g.} $N$ fermions in the singlet sector).
     
Formaly, we can again write eq. (5.31) in the same form as eq. (4.41):
$$
\eqalignno{
Z^{(N)}_R(\beta ) & = \sum_{S} M^{S}_{R\otimes S}
e^{-i\beta \omega \sum^N_{\kappa =1}(\ell _\kappa  + {1 \over 2})} 
& (5.32)
}
$$
with the multiplicity $M^{S}_{R\otimes S}$ given by eq.(4.44).  But
one has now to give a meaning to the imaginary energy levels.
     
For this purpose let us analyze the simplest case of the singlet
representation for which
$$
\eqalign{
M^{S}_{R\otimes S}= 1
}
\eqno{(5.33)}
$$
     
Hence,
$$
\eqalignno{
Z^{(N)}_{{\rm sing}}(\beta ) & = \sum_{\{ \ell _\kappa  > \ell
 _{\kappa +1}\} }
e^{-i\beta \omega \sum^N_{\kappa =1} (\ell _\kappa  + {1 \over 2})} =
 \cr
& = { e^{-i\beta \omega {N^2 \over 2}} \over
(1-e^{-i\beta \omega })(1-e^{-2i\beta \omega }) \cdots
(1-e^{-Ni\beta \omega })} & (5.34)
}
$$
which can be reproduced from the result for the singlet representation
(4.45) of the standard oscillator by change $\omega \rightarrow
i\omega $.
     
How to give a meaning to eq.(5.34)?  Let us pass from the microcanonical
to the grand canonical ensemble.  The singlet partition function takes
the form
$$
\eqalignno{
& Z_{\rm sing} (\mu , \beta ) = \sum ^\infty _{N=0} Z^{(N)}(\beta )
 e^{\beta \mu } = \cr
& = {\rm exp} \sum^\infty _{k=0} log(1 + e^{\beta [\mu -i\omega (k + {1
 \over 2})]})
& (5.35)
}
$$
in the complete analogy with the well-known expression for the grand
canonical partition function of fermions in the oscillatorial
potential.  As usually, the complex energy levels have to be
understood as resonances [1,4-7]:
$$
\eqalignno{
 f_{\rm sing} (\mu ,\beta )&= \log Z_{q\rm sing} (\mu ,\beta )
 =\int^{+\infty} _{-\infty } d\epsilon {1 \over \pi } \sum^\infty
 _{k=0} {k + {1 \over 2} \over ({\epsilon \over \omega })^2 + (k + {1
 \over 2})^2} log [ 1 + e^{\beta (\mu - \epsilon )}] \cr & = \int
 ^\infty _{-\infty } d\epsilon \rho _{\rm sing} (\epsilon )
\log (1 + {\rm exp}\beta (\mu  - \epsilon )) & (5.36)
}
$$
where $\rho _{\rm sing}(\epsilon )$ coincides with eq.(5.20).
     
It is assumed that every integral in the sum is to be calculated as a
pole contribution of the denominator in eq. (5.36) ignoring the
singularities of $log$.  In fact, at the lower limit, the integral in
eq. (5.36) is logarithmically divergent and one has to cut the
integration by some minimal energy $\epsilon _0 (\mid \epsilon _0 \mid
\sim N)$, which is defined by the bottom of the original
potential. Furthermore, the sum over resonances also diverges and needs
to be regularized as well. Nevertheless, we have reproduced the
correct result representing $f_{\rm sing}$ as the partition function
of $N$ fermions in the upside-down oscillatorial potential [2]. Now,
assuming that the general structure of answers is the same for all
representations, we can proceed and get for the adjoint using
eq. (C10) the following formal expression:
$$
\eqalignno{
Z^{(N)}_{\rm adj}(\beta ) = & {e^{-{3 \over 2}i\omega \beta } \over
 1-e^{-i\omega \beta }}
 e^{-i\omega \beta (N-1)}Z^{(N-1)}_{\rm sing}(\beta ) -
 e^{-i\omega \beta N}Z^{(N)}_{\rm sing}(\beta ) &
 (5.37)
}
$$
and, in the grand canonical form,
$$
\eqalignno{
&Z_{\rm adj}(\mu ,\beta )  = \sum^\infty _{N=0}e^{\beta \mu
 N}\ d_{adj} \ Z^{(N)}_{\rm adj}(\beta ) = \cr
& = \bigg( \bigg({1\over\beta}{\partial\over\partial\mu}\bigg)^2 - 1\bigg)
\bigg( A(\mu ,\beta )Z_{\rm sing}(\mu  - i\omega , \beta ) + B(\mu ,\beta
 )\bigg) & (5.38)
}
$$
where we have represented the dimension of the adjoint ,$d_{adj} =
N^2-1$, as the differential operator and pulled it out of the sum over
$N$.
$$
\eqalign{
A(\mu ,\beta ) = {{\rm e}^{ \beta (\mu - {3 \over 2}i \omega) }
 \over
1 - {\rm e}^{-i\beta \omega }} - 1
}
\eqno{(5.39)}
$$
If we define that
$$
\eqalign{
Z_{\rm adj}^{(0)}(\beta ) = Z_{\rm sing}^{(0)}(\beta ) \lim_{N\to 0}
{e^{i\omega \beta } - e^{-i\omega \beta N} \over 1 - e^{-i\omega \beta
 }}  = -1
}
\eqno{(5.40)}
$$
then
$$
\eqalign{
B(\mu ,\beta ) = 0
}
\eqno{(5.41)}
$$
In any case, this constant is determined by the first terms in the
grand canonical expansion in eq. (5.38) and cannot influence universal
properties of $Z_{\rm adj}(\mu ,\beta )$.
     
Now using the same procedure which has led us from eq. (5.35) to
 eq. (5.36) we obtain:
$$
\eqalignno{
& Z_{\rm adj}(\mu ,\beta ) = \bigg( \bigg({1\over\beta}{\partial\over\partial\mu}
\bigg)^2 -1 \bigg) A(\mu ,\beta )Z_{\rm sing}(\mu -i\omega
 ,\beta ) = \cr
& = \bigg( \bigg({1\over\beta}{\partial\over\partial\mu}\bigg)^2 -1 \bigg) 
A(\mu ,\beta ) {\rm exp}\sum^\infty _{k=1} \int^{+\infty}_{-\infty }{d\epsilon
 \over \pi\omega }
 {k + {1 \over 2} \over ({\epsilon \over \omega} )^2 +
(k + {1 \over 2})^2} \log \big( 1 + e^{\beta (\mu  - \epsilon )}\big)
 \cr
& = \bigg( \bigg({1\over\beta}{\partial\over\partial\mu}\bigg)^2 -1 \bigg) 
A(\mu ,\beta ) {\rm exp}\int ^\infty _{-\infty } d\epsilon \rho
 _{\rm adj}(\epsilon )
\log \big( 1 + e^{\beta (\mu  - \epsilon )} \big) & (5.42)
}
$$
where
$$
\eqalign{
\rho _{\rm adj}(\epsilon )  
  = -{1 \over \pi\omega } Re \psi (i{\epsilon  \over \omega } + {3 \over 2})
  + {1 \over \pi\omega } \log \Lambda 
}
\eqno(5.43)
$$
or
$$
\eqalign{
\rho _{\rm adj}(\epsilon ) = \rho _{\rm sing}(\epsilon ) - {1 \over
 2\pi \omega}
{1 \over ({\epsilon  \over \omega })^2 + {1 \over 4}}
}
\eqno{(5.44)}
$$
     
We see that the effective density of states in the adjoint
 representation differs from the singlet one by one missing Lorentzian
 (resonance).
     
The factor $A(\mu ,\beta )$ could be rewritten with the same
 argumentation in the form
$$
\eqalign{
A(\mu ,\beta ) = \int ^\infty _{-\Lambda^2} d\epsilon\ 
 \rho _{\rm adj}(\epsilon)\ e^{\beta (\mu -\epsilon )} - 1
}
\eqno{(5.45)}
$$
and interpreted as the partition function of one particle decoupled
from the others, even though we do not have any microscopic arguments
supporting this interpretation.  Nevertheless, it seems to give just a
non-universal factor of the order $\exp (\beta \Lambda^2)$ which does
not affect universal behavior. Of course, this particle does interact
with others at distances of the order $\Lambda$, but, being taken into
account, this interaction produces only exponentially small in $N$
terms and does not change the double scaling limit.
     
Unfortunately, except for the adjoint representation, the procedure of
the analytical continuation is ambiguous.  If we collect together four
representations of the next order (corresponding polynomials given by
eqs.(C11), (C12), (C13)) and implement the same procedure as led us to
eq. (5.38), we shall get
\def\dm{{1\over\beta}{\partial\over\partial\mu}}
\def\ob{i\omega\beta}
$$
\eqalignno{
 Z_2 (\mu ,\beta )&  =
\sum_N\ e^{\beta\mu N} \big( 2d_{A_2} Z_{A_2}^{(N)}(\beta)
           +  d_{B_2} Z_{B_2}^{(N)}(\beta)
           +  d_{C_2} Z_{C_2}^{(N)}(\beta)\big)\cr
&={e^{\beta(2\mu-2i\omega)}\over(1-e^{-\ob})(1-e^{-2\ob})}
\bigg\{ {e^{-\ob}\over2}\bigg(\bigg(\dm\bigg)^2 - 1\bigg)
\bigg(\bigg(\dm\bigg)^2 - 4 \bigg) +\cr
&{e^{-2\ob}\over4}\bigg(\dm\bigg)^2 \bigg(\dm-1\bigg)\bigg(\dm+3\bigg)+\cr
&{1\over4}\bigg(\dm\bigg)^2 \bigg(\dm+1\bigg)\bigg(\dm-3\bigg) \bigg\}
Z_{\rm sing}(\mu - 2i\omega,\beta)\cr
&+{e^{\beta(\mu-{1\over2}i\omega)}\over(1-e^{-\ob})}
\bigg\{ {1\over2}\bigg(\bigg(\dm\bigg)^2 - 1\bigg)
\bigg(\bigg(\dm\bigg)^2 - 4 \bigg) +\cr
&{1\over4}\bigg(\dm\bigg)^2 \bigg(\dm+1\bigg)\bigg(\dm-3\bigg) \bigg\}
Z_{\rm sing}(\mu - i\omega,\beta)\cr
&+{1\over2}\bigg(\bigg(\dm\bigg)^2 - 1\bigg)
\bigg(\bigg(\dm\bigg)^2 - 4 \bigg) 
Z_{\rm sing}(\mu ,\beta) &(5.46)
}
$$
where we have substituted $d_{A_2}={1\over4}(N^2-1)(N^2-4)$;
$d_{B_2}={1\over4}N^2(N-1)(N+3)$; $d_{C_2}={1\over4}N^2(N+1)(N-3)$ and
converted them into the differential operators.

Eq. (5.46) should contain, as we believe, contributions describing
different configurations having the total vortex (anti-vortex) charge
equal to 2.  Moreover, there should be a contribution from two vortex
pairs living on different surfaces (since we deal with the partition
function rather than the free energy). In order to perform an
analytical continuation, one should presumably separate all those
terms and treat them differently.
     
\chapter{6. Physical results  in the double scaling limit}

The results for the upside down oscillator given by eqs.(5.36) and
(5.42) allow us to obtain the double scaling limit for the
compactified one dimensional string (or for the XY-model coupled to
gravity) in the sector with one vortex-anti-vortex pair.
     
The partition function for the singlet representation eq. (5.36) was
investigated in ref. [2]. It corresponds to the case of no vortices,
when only nontrivial cycles of a two-dimensional manifold (given by a
Feynman graph with non-trivial topology) can wrap around the time
circle of length $\beta$.

{\it Singlet partition function  }

Let us repeat the calculation of $f_{\rm sing} (\beta,\mu)$ from
ref. [2] for the consistency of content.  We have to remember that
the eq.(5.36) describes a filling by free fermions of both sides of
the inverted oscillatorial potential. However, in the string
perturbation theory which will correspond here to $1/\mu$-expansion we
have to take into account only the states on one side of this
potential, which corresponds to the perturbative vacuum of this string
theory. This is reflected by an overall factor $1/2$ in the following
formula for the singlet free energy:
$$
\eqalign{
 f^{(pert)}_{\rm sing}(\mu) =f_{sing}^{reg}(\mu)+
{1\over 2} \int ^\infty _{-\infty } d\epsilon \rho _{\rm sing}
 (\epsilon )
\log \left(1 + e^{\beta (\mu- \epsilon  )}\right)
}
\eqno{(6.1)}
$$
where $f_{sing}^{reg}(\mu)=A_0(\lambda)+A_1(\lambda)\mu
+A_2(\lambda)\mu^2+\cdots$ is the regular part of expansion of the
grand canonical free energy in $\mu$.  Using an integral
representation for $\rho _{\rm sing}(\epsilon )$
$$
\eqalign{
\rho _{sing}(\epsilon ) = 
{1 \over 2\pi } Re \int ^\infty _{\Lambda ^{-1}} dx {e^{i\epsilon x}
\over {\rm sinh} { x\over 2}} }
$$
(we put $\omega = 1$, fixing appropriately the scale of $\epsilon $)
  and integrating over $\epsilon $ in the eq. (6.1), we obtain the
  following formula for the free enery of singlet states
$$
\eqalign{
{\partial ^3 f _{\rm sing} \over \partial \mu^3} = {\beta \over 2\pi} 
Im \int ^\infty _0 dx 
{{x\over 2} \over {\rm sinh} {x \over 2 }}
{{\pi x \over \beta }\over {\rm sinh} {\pi x \over  \beta }} e^{i\mu x}
}
\eqno{(6.2)}
$$
     Expanding $f _{\rm sing}$ in $1/\mu$ and introducing a cut-off
$\Lambda$ to regularize the logarithmic divergences we get:
$$
\eqalign{
f^{(sing)}(\mu) = {\beta \over 2\pi} \bigg[ -{1\over 2}\mu^2 \log
(\mu/\Lambda^2) - {1\over
24}\big(1+\big({2\pi\over\beta}\big)^2\big)\log (\mu/\Lambda^2)
+\sum_{k=2}^\infty \mu^{-2(k-1)} f_k(\beta)
\bigg]}
\eqno{(6.3)}
$$
where we introduced  polynomials in ${1\over\beta}$
$$
\eqalign{ f_k(\beta)=
(2k-3)! \ \ 2^{-2k} \sum_{n=0}^{k} 
\big({2\pi\over \beta}\big)^{2n}
{(2^{2(k-n)}-2)(2^{2n}-2)\vert B_{2(k-n)}\vert\vert B_{2n}\vert \over
[2(k-n)]![2n]!}}
$$
and $B_m$ are the Bernoulli numbers.

We do not have a direct interpretation of this grand canonical
partition function as a sum over world sheets embedded into a time
circle and classified by specific topologies. Such interpretation
exists only for the canonical ensemble where we fix the variable $N$
instead of $\mu$. The relation between two ensembles is given by the
integral transform:
$$
\eqalign{  \exp[f_N^{sing}]=\oint d\mu  \exp[-\beta \mu N+ 
f^{sing}(\mu)] 
}
\eqno{(6.4)}
$$
where the integration contour encircles the point $e^{\beta
\mu}=0$. The $1/N$ expansion of $f_N^{sing}(\lambda)$
provides the string partition functions of given genera.

In the double scaling limit which is the continuum limit of the string
theory we have to extract only the lading singularity at $\lambda\to
\lambda_c$ 
\foot{ $\lambda_c$ is defined in such a way 
that in the planar limit $N\to\infty$ when the temperature
$\beta^{-1}$ is effectively zero (the spherical string amplitude does
not feel the compactness of ``time'') we have $\mu(\lambda_c)=0$}. 
To do it let us rewrite (6.4) in the form
$$
\eqalign{  \exp[f_N^{sing}]=\oint d\mu  
\exp[{\beta\over 2\pi}\mu \mu_0\log\mu_0 + f^{sing}(\mu)]  } 
\eqno{(6.5)}
$$ 
where we introduced the parameter $\mu_0$ by the formula 
$$
 \mu_0\log(\mu_0/\Lambda^2)=2\pi[A_1(\lambda)-N]\sim N \Delta
$$
with $\Delta=(\lambda^2_c - \lambda ^2)/\lambda^2_c.$
 
The integral can be calculated in terms of the series in $1/\mu_0$
which is equivalent to the $1/N$ expansion. To generate this series we
have to apply the saddle point approximation and then to take
systematically into account the fluctuations around the saddle
point. The saddle point for $\mu(\mu_0)$ can be found from the
equation:
$$
\eqalign{
\mu_0\log(\mu_0/\Lambda^2) =
{2\pi\over \beta}{\partial f^{pert}_{sing}(\mu)\over\partial\mu}  } 
\eqno{(6.6)}
$$
Solving it iteratively we obtain:
$$
\eqalign{ \mu=\mu_0\bigg[1-{1\over \log(\mu_0/\Lambda^2)}
\left( \sum_{k=1}^\infty \mu_0^{-2k} 2(k-1) f_k(\beta)\right)+ 
O({1\over \log^2(\mu_0/\Lambda^2)}) \bigg]   } 
\eqno{(6.7)}
$$
We see that in the double scaling limit when $\mu_0$ is kept finite
and  $\Lambda\to \infty$ we can take $\mu=\mu_0$ as a saddle point.

 Then we have to put $\mu=\mu(\mu_0)+\delta\mu$ and integrate over
$\delta\mu$ in the Gaussian approximation.  The contribution to
$f_N^{sing}$ from this Gaussian integration  looks as $-{1\over
2}\log\log(\mu_0/\Lambda^2)+ O({1\over \log(\mu_0/\Lambda^2)})$. These
terms should be neglected with the same accuracy.

 The net result of this calculation is that with the accuracy up to
inverse logarithmic corrections we substitute the integral (6.5) by
its saddle point value at $\mu=\mu_0$ and get the following result for
the canonical free energy of singlet states [2]:
$$
\eqalign{
f_N^{sing} = {\beta \over 2\pi} 
\bigg[ {1\over 2}\mu_0^2 \log (\mu_0/\Lambda^2) -
{1\over 24}\big(1+\big({2\pi\over\beta}\big)^2\big)\log (\mu_0/\Lambda^2) 
+\sum_{k=2}^\infty \mu_0^{-2(k-1)} f_k(\beta)
\bigg]}
\eqno{(6.8)}
$$
Note the change of sign of the first term (spherical free energy) with
respect to the similar term in (6.3).

As it was noticed in ref. [2] eqs. (6.2),(6.3) and (6.8) possess a
T-duality symmetry with respect to the inversion of the radius of
compactification:
$$
\eqalign{
{\beta  \over 2\pi } \rightarrow {2\pi  \over \beta } ; \quad
\mu   \rightarrow { \beta  \over 2\pi } \mu}
$$     
This important property remains to be true in every order of the
topological (1/N) expansion (we recall that $\mu \sim N {\Delta \over
\vert \log \Delta \vert}$ in the planar limit).

\vskip2cm

{\it Adjoint partition function }

The analytical continuation of the partition function for the adjoint
representation (eq. (5.42)) looks ambiguous but nevertheless we assume
that its universal part can be written as follows

$$
\eqalign{
Z_{adj} = \big({1\over\beta^2}{\partial^2\over \partial \mu^2}-1\big)
\exp f_{adj}(\mu,\beta)
}
\eqno{(6.9)}
$$
\noindent
where $f_{adj}(\mu,\beta)$ is determined similarly to
$f_{sing}(\mu,\beta)$ but with the density of states given by
eq. (5.45)
$$
\eqalign{
\rho _{\rm adj}(\epsilon ) = \rho _{\rm sing}(\epsilon ) - 
{1 \over \pi } {1/2 \over
 \epsilon ^2 + 1/4}
}
\eqno{(6.10)}
$$
Hence, we have
$$
\eqalign{
\delta = f_{\rm adj} -f_{\rm sing} =
{1\over 2} \int d\epsilon 
(\rho _{\rm adj}(\epsilon ) - \rho _{\rm sing}(\epsilon ))
\log\big(1 + e^{\beta(\mu - \epsilon)}\big)}
\eqno(6.11)
$$
Using the representation 
$$
\eqalign{
{z\over \epsilon^2 + z^2} = Re \int_0^\infty dx e^{ix\epsilon-zx}
}
\eqno(6.12)
$$
and integrating in eq. (6.11) over $\epsilon$ we obtain for the third
derivative of $\delta$ the following representation
$$
\eqalign{
{\partial \delta \over \partial \mu} = {1 \over 2 }\int_0^\infty dx
\ \ {{\rm sin}(\mu x) \ \  {\rm sinh}{x\over 2}\over
{\rm sinh}{\pi x \over\beta}}
}
\eqno(6.13)
$$
And the topological expansion of $\delta$ takes the form
$$
\eqalignno{
\delta(\mu) &= {\beta \over 2\pi} \bigg[ \log (\mu/\Lambda^2)
- {1\over 8\mu^2}\big(1-{1\over 3}\big({2\pi\over\beta}\big)^2\big)
 \cr &- \sum_{m=2}^\infty \mu^{-2m}(-1)^m (2m-1)! \ 2^{-2m}
 \sum_{n=0}^{m} \big({2\pi\over \beta}\big)^{2n}
 {(2^{2n}-2)B_{2n}\over [2(m-n)]!!(2n)!}\bigg]&(6.14)}
$$
   
From eq. (6.14) we conclude that
$$
\eqalignno{
& f_{\rm adj} - f_{\rm sing} \underrel{\sim}{\mu\to\infty}
 {\beta  \over 2\pi }  \log{ \mu \over \Lambda ^2}
&(6.15)
}
$$
     
Since $\mu \sim \Delta\cdot N$ in the first approximation (and
$\Lambda^2\sim N$), we reproduce in eq. (6.15) the result (1.1) from
the paper [3]: the spectrum of angular (vortex) excitations is
separated from the singlet (vortex-free) spectrum by a logarithmically
big gap.  Note that the expression (6.14) is not self-T-dual anymore.
     
This fact can be qualitatively explained from the comparison of the
sum-over-surfaces picture for the correlator eq. (4.20) propagating
the singlet states and the correlator (4.28) propagating the adjoint
states.  The first one is given by the sum over all Feynman graphs
with two vertices fixed (on a graph and in the target space).  These
two vertices are arbitrary and can belong to any two different loops
or the same loop on a graph.
     
In the case of the correlator (4.28) one has to pick up two vertices
belonging to the same loop on a graph, which is only a part of
(positively defined) contributions of the correlator (4.20).  Hence
$$
\eqalign{
{\partial ^2 \over \partial \alpha _1 \partial \alpha _2} D_{\alpha
 _1\alpha _2}
(\beta ) \mid_{\alpha _1=\alpha _2=0} = D_0(\beta ) > K_0(\beta )
}
\eqno{(6.16)}
$$
If we take for the definition of the gap the large $\beta $
 asymptotics:
$$
\eqalign{
D_0(\beta ) \sim  {\rm exp}(-m_{\rm sing} \cdot \beta )
}
\eqno{(6.17)}
$$
$$
\eqalign{
K_0(\beta ) \sim {\rm exp}(-m_{\rm adj} \cdot \beta )
}
\eqno{(6.18)}
$$
we come from eq. (6.16) to the conclusion that
$$
\eqalign{
m_{\rm adj} > m_{\rm sing}
}
\eqno{(6.19)}
$$
     
Hence, it is not surprising that
$$
\eqalign{
m_{\rm sing} {\sim } {1 \over \vert \log \Delta\vert }
\underrel {\rightarrow} {\Delta \to 0}0 }
\eqno{(6.20)}
$$
and
$$
\eqalign{
m_{\rm adj} {\sim }
\vert \log \Delta \vert  \underrel{\rightarrow} {\Delta \to 0}+\infty
}
\eqno{(6.21)}
$$
     
As we demonstrated in the section 4, higher representations correspond
to the multi-loop amplitudes of open strings with free boundary
conditions on the edges of the world sheet. We have no parameter (like
the mass of a particle at the end-points of the open string) to adjust
in order renormalize the ``boundary tension'' and to make the
characteristic size of the boundary macroscopic.  Eqs. (6.20) and
(6.21) show that the characteristic boundaries have very short
lengths, and the contribution of few vortices is negligible in
comparison with the contributions of vortex-free configurations.
     
However, the vortices have a considerable entropy, which causes at
 some $\beta $ the \break Berezinski-Kosterlitz-Thouless phase
 transition [13,14]. Knowing the partition function for one
 vortex-anti-vortex pair, we can calculate the critical value,
 $\beta_{KT}$, as follows. From the eqs. (3.15), (6.8) and (6.15) we
 obtain in the spherical approximation in the canonical (fixed $N$
 ensemble) the following partition function of the vortex anti-vortex
 pair in the $c=1$ string theory [3]:
 $$
\eqalignno{   {\cal Z}_N(\beta,\lambda)&\simeq Z_{sing}\left(1+ N^2
 e^{f_{adj}-f_{sing}}+\cdots\right) \cr
&\simeq \exp\left({\beta\over 2\pi}
 N^2{1\over 2}\Delta^2[\log(\Delta)+{\it const}\Delta^{{\beta\over
 2\pi}-2}+\cdots]\right)    
 & (6.22)   }   
$$
We see that the second term (as well as the contributions with more
vortices and anti-vortices) is irrelevant with respect to the first
one, describing the usual scaling of the $c=1$ string with no
vortices, when $\beta>4\pi$. However, it becomes more important than
the first term for $\beta<4\pi$ and the dilute gas approximation for
vortex-antivotex pairs used in the eq. (6.22) is not valid any more.
The world sheet of the string will be immediately densely populated by
the strongly interacting vortex plasma. It is a typical picture of the
Berezinski-Kosterlitz-Touless phase transition.

This phase transition occurs in our model at the same inverse
temperature
$$
\eqalign{
\beta_{KT} = 4\pi
}
\eqno{(6.23)}
$$
as in the model of planar rotors ($XY$ model) on the plane. Indeed,
the answer (6.22) should be compared with the configuration integral
of the vortex-anti-vortex pair on the plane [13,14] (in the same
normalization):
$$
\eqalign{
Q = \int d^2x \int d^2y \exp - { \beta\over 2\pi }\log {\vert x-y\vert^2
\over r_o^2}
\sim V^{2 - { \beta\over 2\pi }}
}
\eqno{(6.24)}
$$
where $V$ is the volume  of the space.

\chapter{7. A possible approach to  D+1-dimensional bosonic string}
     
A natural generalization of the matrix model describing
one-dimensional strings to the case of D+1-dimensional bosonic strings
is the following D+1-dimensional scalar matrix field theory:
$$
\eqalign{
Z_N(\beta ,\lambda ) = \int{\cal D}^{N^2}\varphi (x,t) {\rm exp} - Ntr
\int dt \int d^Dx \big[ {1 \over 2}(\partial ^\mu \varphi \partial
_\mu \varphi ) + V(\varphi )\big] }
\eqno{(7.1)}
$$
with the potential $V(\varphi )$ given by eq.(4.3) or of a more general
form.
     
The analogy with the string theory is again based on the Feynman
graphs: we obtain for the free energy $f_N^{(D)} = { 1 \over N^2} log
Z_N^{(D)}$ the following diagrammatic expansion
$$
\eqalign{
f_N(\beta ) = \sum^\infty _{g=0}N^{-2g} \sum^\infty _{k=0}\lambda ^k
\sum_{G_g^{(n)}} \int d^{D+1}x_1 \cdots d^{D+1}x_n\prod_{<ij>\in
 G_g^{(n)}}
D(x_i - x_j)
}
\eqno{(7.2)}
$$
where
$$
\eqalign{
D(x) = {1 \over (2\pi )^{D+1}}\int d^{D+1}p\ e^{i(px) }{1\over p^2 +
\omega ^2} }
\eqno{(7.3)}
$$
     
As long as we have the superconvergibility of the integrals in
eq. (7.2) (it is true, say, for $D < D_c = 5$ for $\varphi ^3$ graphs)
we may hope that the particular microscopic definition of $D(x)$ is
irrelevant, and for big enough k's in eq. (7.2) we have defined the
$D+1$-dimensional bosonic string field theory.
     
Unfortunately, the theory (7.1) is much more complicated than the
theory (2.1).  We cannot, for example, obtain a representation similar
to the Hamiltonians (3.16) for the eigenvalues of the matrix field:
the corresponding ``connection'' $A_\mu = \Omega ^+\partial _\mu
\Omega $ (compare with eq.(3.3)), obeys now the more complicated
constraint, $\partial _\mu A_\nu - \partial _\nu A_\mu + [A_\mu, A_\nu
] = 0$, and the integration over $A_\mu $ is highly nontrivial.  The
importance of the angular variables for $C=D+1>1$ physics is reflected
in the fact that the eigenvalues may be now not the only and even not
the most important degrees of freedom.
     
On the other hand, if we still hope that the physics at $C>1$ bears
some universal features, we may admit that the particular shape of the
potential $V(\varphi )$ should not be important.  All the results for
$C<1$ show that only singular points like local maximum or inflection
points are essential for the continuum limit of the corresponding
string theory, and all the rest in the shape of the potential after an
appropriate rescaling serves as a cut-off.
     
This brings to mind the idea that it may be sufficient to consider the
most generic unstable potential:
$$
\eqalign{
V(\varphi ) = -{m^2 \over 2} \varphi ^2
}
\eqno{(7.4)}
$$
Forgetting for a while about the rest of the potential, we can use the
solvable Gaussian theory
$$
\eqalign{
Z_N = \int {\cal D}^{N^2}\varphi (x) {\rm exp} - {N \over 2} tr
\int d^{D+1} x \big[ (\partial _\mu  \varphi )^2 + m^2 \varphi ^2\big]
}
\eqno{(7.5)}
$$
in order to get the results for the unstable theory (7.4) by means of
the analytical continuation
$$
\eqalign{
m \rightarrow im
}
\eqno{(7.6)}
$$
and an appropriate introduction of a cut-off.  If we will be unable to
find this procedure so the whole approach based on the matrix model
(7.1) does not describe any universal bosonic string theory for $c >
1$.
     
Again as in the case $c=1$, the Gaussian theory (7.5) is too poor to
be a starting point for an analytical continuation, and we have to
consider the states obeying some particular representation of the U(N)
group (since the potential $trV(\varphi )$ is still invariant under
$\varphi \rightarrow \Omega ^+\varphi \Omega $ transformation and so
is the cut-off procedure).
     
It is natural to hope that the ground state of the system is again the
singlet under the group transformations; therefore, we have to extract
only the singlet part of eq. (7.5) and then to perform the analytical
continuation.
     
In order to find the partition function for the singlet states we
introduce again the twisted partition function: we compactify the time
dimension on the circle of a length $\beta $ ($t \in (0, \beta)$
again) and for the remaining $D$ directions we use the momentum
representation:
$$
\eqalignno{
Z_N^D(\beta ,\Omega ) & = \int {\cal D}^{N^2}\varphi_p(t)\ {\rm exp}{N
 \over 2}tr
\int ^\beta _0 dt \int d^{D}p\big[ \dot{\varphi }^2 - (p^2 + m^2)\varphi^2
 \big] \cr
& \varphi_p(0) = \Omega ^+\varphi_p (\beta )\Omega  & (7.7)
}
$$
where we have introduced the U(N) twisted periodic initial conditions.
     
For fixed p every matrix element can be considered as an independent
 oscillator with the effective frequency
$$
\eqalign{
\omega_p  = \sqrt{p^2 + m^2}
}
\eqno{(7.8)}
$$
Hence, in the complete analogy with eqs. (4.33) or (5.30) we can
write:
$$
\eqalign{
Z_N^{(D)}(\beta ,\theta ) = {\rm exp} - \sum^N_{k\geq j=1} F(\beta
,\theta _k-\theta _j) }
\eqno{(7.9)}
$$
where
$$
\eqalign{
F(\beta ,\theta ) = L^D\int d^D p \log \big[ cosh {\beta  \over
 2}
\sqrt{p^2 + m^2} - cos \theta \big]
}
\eqno{(7.8)}
$$
     
L is an infrared cut-off (the size of the box where the system is put
in the D dimensional target space), and the twist matrix $\Omega $ is
chosen in the diagonal form
$$
\eqalign{
\Omega  = diag (e^{i\theta _1}, e^{i\theta _2}, \cdots e^{i\theta _N})
}
\eqno{(7.9)}
$$
     
Then the recipe to find the ground state $E_0$ of our model is to
calculate the singlet partition function:
$$
\eqalign{
Z_{\rm sing}^{(D)}(\beta ) =
\int ^{2\pi }_0 \cdots \int ^{2\pi }_0 \prod ^N_{k=1}
{d\theta_k \over 2\pi }
\mid \Delta (e^{i\theta }) \mid ^2 Z^{(D)}_N(\beta ,\theta )
}
\eqno{(7.10)}
$$
to perform the analytical continuation (7.6), introducing the cut-off
on the way, which should defined the dependence on the cosmological
constant, and to find the asymptotics
$$
\eqalign{
Z_{\rm sing}^{(D)}(\beta ) \underrel { \longrightarrow }
{\beta  \rightarrow \infty } e^{-\beta E_0}
}
\eqno{(7.11)}
$$
     
Of course, all these steps do not seem to be trivial. One of many
obstacles to overcome is the highly divergent integral in the
eq. (7.8). one has to learn how to deal with these divergences.  But
what is hopeful here is the fact that the integral in eq. (7.10) goes
over only N (and not $N^2$) variables and is in principle of a saddle
point type in the large N limit.

Another way to proceed is to try to use eq. (4.39) in order to obtain
a representation similar to eq. (4.41):
$$
\eqalignno{
Z^{(D)}_R(\beta) &= \int_0^{2\pi} \prod^N_{k=1} {d\theta_k \over 2\pi
 } \mid \Delta (e^{i\theta }) \mid ^2 \hchi _R (e^{i\theta })
\prod_{\{p\}} \prod_{k\leq j} {\rm e}^{F(\beta,\theta_k - \theta_j)}\cr
&= \sum_{S_p}{\rm e}^{-\beta\omega_p\sum_{k=1}^N(\ell_{k,p}+{1\over2})}
\int (dU)\hchi_R(U)\prod_{\{p\}}\hchi_{S_p}(U)\overline{\hchi}_{S_p}(U)\cr
&=\sum_{S_p}M_{R[\otimes S_p]}^{[\otimes S_p]}{\rm e}
^{-\beta\omega_p\sum_{k=1}^N(\ell_{k,p}+{1\over2})}
&(7.12)}
$$
where $\ell_{k,p}$ are defined analogously to eq. (4.43) and attached
to each point of the $p$ space. $\sum_{S_p}$ is the sum over all
representations at all points. In the continuum limit it becomes a
fermionic path integral.  $M_{R[\otimes S_p]}^{[\otimes S_p]}$ plays
the role analogous to the corresponding factor in eq. (4.41) making
the system highly non-trivial. It gives rise to the same selection
rule (3.14) for allowed representations $R$ as in the one-dimensional
case.

  We hope to come back to this problem in the future and in this paper
we propose eqs. (7.9), (7.10) as a definition of a D+1-dimensional
bosonic string theory directly in the continuous limit, as it was
discussed in the introduction.
     
Another interesting question is the role of D-dimensional partition
functions in higher representations.  They might describe vortex-like
excitations (i.e. monopole) which might be quite important for the
physics of strings.  It is not clear whether they are separated by a
gap from the singlet vacuum.

\chapter{8. Discussion.}

The aim of this paper was to demonstrate the possibility to extract
physical information about $1D$ bosonic string by means of the
analytical continuation of quantities known for the standard matrix
oscillator. It was shown that not only the singlet (with respect to
the $U(N)$ symmetry of the model) but also the adjoint partition
functions can be obtained in the double scaling limit with the help of
this procedure. It was shown that the adjoint partition function
describes the one-vortex-anti-vortex-pair sector of the $XY$-model
coupled to gravity.  It enabled us to calculate the critical
temperature for the Berezinski-Kosterlitz-Thouless phase transition.

However, the method of analytical continuation obviously suffers from
ambiguities which appear because the high symmetry of the matrix
oscillator cannot be preserved in the upside-down case. Any cut-off
imposed in order to stabilize the system decreases this
symmetry. Therefore we had to guess a dependence on the cut-off using
close analogy between the singlet and adjoint partition
functions. However, in the case of higher representations, one
apparently needs some microscopic information in order to make the
continuation unambiguous. To begin with, one needs {\sl e.g.}  the
decomposition of the sum over all representations with respect to the
contributions of particular vortex-anti-vortex combinations. It may be
done matching the partition functions with propagators of non-singlet
states as it was done in this paper for the adjoint
representation. Another useful trick would be to introduce a
regularization similar to one proposed in ref. [11] in the singlet
case.

It is tempting to interpret the non-local matrix variable
$$
\eqalign{
\Omega(0,\beta) = \widehat T {\rm exp}~ i\int^\beta _0 dt A(t)
}
\eqno{(8.1)}
$$
as an operator creating a vortex of the unit charge. It would be
interesting to include the angular degrees of freedom in the
collective field representation [22] as well as eigenvalues.

Although, knowing the partition function for one vortex-anti-vortex
pair, one can correctly estimate the position of the
Berezinski-Kosterlitz-Thouless phase transition, in order to get a
solution at the critical point, one should sum over all or, maybe,
over the most important representations. This problem is far from
being solved. However, it seems that properties of the $XY$-model
interacting with 2D quantum gravity are quite similar to those of the
$XY$-model on the plane: for $\beta > 4\pi$ one has the
dilute-gas-of-dipoles phase, where a vortex and an anti-vortex are tied
together forming a dipole; and for $\beta < 4\pi$ one has the plasma
phase, where interactions between vortices are screened and an
emerging finite correlation length leads to the change of the matter
central charge from $c = 1$ to $c=0$.

The model under investigation is the first example of a matrix model
where the angular degrees of freedom (and not just the eigenvalues) of
the matrix field play an important role.  Now, it is clear that the
angular degrees of freedom describe vortices and hence they are of a
great concern in many physical applications. They become crucial when
the central charge is bigger than 1. Apparently, one has to pay much
more attention to them in order to understand the nature of this phase
of the string theory from the matrix model point of view.

\chapter{Aknowledgements}

The authors are very indebted to E.Brezin and I.K.Kostov for many
useful discussions and inspiring proposals on the early stage of this
work.  We also thank L.Alvarez-Gaume, J.-M.Daul, D.Gross, V.Fateev,
A.Kirillov, I.Klebanov, A.Migdal, F.Smirnov, A.B.Zamolodchikov,
Al.B.Zamolodchikov and especially M.Douglas for many valuable
comments.

\vfill

\appendix{A}{}
     
In this appendix we derive the selection rule (3.14) for
representation contributing to the sum (3.15) and discuss the behavior
of eigenfunctions of the Hamiltonian (3.16) with respect to
permutations of the coordinates $z_i$. The Fourier decomposition of
$\Psi$-functions of the original matrix Schroedinger equation gives
rise to the expansion of the Green function (defined in the case of
the harmonical oscillator by eqs.  (4.14) and (4.15))
\eqn\GXY{
G(X,Y) = \sum_{\{ R \} } \sum^{d_R}_{a,b=1}g^R_{ab}(\lambda ,\mu )
D^R_{ab}(U)     }
where $\lambda$ and $\mu$ are eigenvalues of N$\times$N matrices $X$ and
$Y$ ( $X =\Omega _1\lambda \Omega _1^+, Y =\Omega _2\mu \Omega _2^+$)
 and $U = \Omega _1^+\Omega _2 \in $ U(N); $D^R_{ab}(U)$ is a matrix
element in the space of a representation $R$. The symmetry
\eqn\GXOYO{\eqalign{
G(X, \Omega Y \Omega ^+) = G(\Omega ^+X \Omega , Y)
}}
means that eq.(A1) is invariant under left and right shifts of U by a
diagonal unitary matrices: 
$U_{kl} \to U_{kl} e^{i(\theta _k - \varphi_l)}$. By definition,
\eqn\GRAB{
 g^R_{ab}(\lambda ,\mu ) = \int (dU) G(\lambda ,U\mu U^+){\bar
 D}_{ab}^R(U) }  
and, from the invariance of the Haar measure under this symmetry,
\eqn\DBB{\eqalign{
g^R_{ab}(\lambda ,\mu ) &= \int (dU)G(\lambda ,U\mu U^+){\bar D}^R_{a'b'}(U)
 \ \prod ^N_k {d\theta _k \over 2\pi }{\bar D}_{aa'}(e^{i\theta })
\int ^{2\varphi}_0 \prod ^N_k {d\varphi _k \over 2\pi }
{\bar D}_{b'b}(e^{-i\varphi })   \cr 
&=  P_{aa'}^R g^R_{a'b'}(\lambda ,\mu )P_{b'b}^R
}}
where $P^R_{ab} = \int ^{2\pi }_0 \prod ^N_{k=1} {d\theta_k \over 2\pi }
{\bar D}_{ab}(e^{i\theta })$ is a projector , since
\eqn\PRSQ{\eqalign{
 (P^R)^2_{ab} &= \int ^\pi _0 \prod _k {d\theta _k \over 2\pi } \int
 ^{2\pi }_0
\prod _k {d\varphi _k \over 2\pi } D_{aa'}(e^{i\theta })D_{a'b}(e^{i\varphi
 })  \cr
&=\int ^{2\pi }_0 \prod _k {d\theta _k \over 2\pi } \int ^{2\pi }_0
\prod _k {d\varphi _k \over 2\pi } 
 D_{ab}(e^{i(\theta +\varphi )}) = P_{ab}^R 
}}
The dimension of the subspace it projects on to is equal to
     
\eqn\DOR{\eqalign{
d^{(0)}_R = {\rm tr} _R P^R = \int ^{2\pi }_0 \prod ^N_{k=1}{d\theta
 _k \over 2\pi }
\hchi _R(e^{i\theta })
}}
Under permutations of $\lambda (\lambda  \rightarrow \rho \lambda \rho
 ^{-1})$ the Fourier
coefficients behave as follows
\eqn\RHO{\eqalign{
 g^R_{ab}(\rho \lambda \rho ^{-1}, \mu ) &= \int (dU)G(U^+\rho \lambda
 \rho ^{-1}U,\mu ) {\bar D}_{ab}^R(U) \cr &= {\bar D}^R_{aa'}(\rho
 )g^R_{a'b}(\lambda ,\mu ) }} and analogously for $\mu $
\eqn\GRBL{\eqalign{
& g_{ab}^R(\lambda ,\rho \mu \rho ^{-1}) = \int (dU)G(\lambda ,U\rho
\mu \rho ^{-1}U^+) {\bar D}^R_{ab}(U) =  \cr &
g^R_{ab'}(\lambda ,\mu ) {\bar D}^R_{b'b}(\rho ^{-1}) =
D^R_{bb'}(\rho) g^R_{ab'}(\lambda ,\mu ) }} 
It is clear from eqs. (A7) and (A8) that $g_{ab}^R(\lambda ,\mu )$ is
a tensor operator with respect to the permutations, i.e. the Weyl
group. It means that properties of the Green function (or the
$\Psi$-functions) with respect to the permutations are determined by
the matrix elements of the Weyl group, which are non-trivial in
general. For self-consistency the subspace defined by eqs.(A4) - (A6)
has to be invariant under all permutations of $\lambda$ and $\mu$.  It
is not difficult to show that it is the subspace of all zero-weight
vectors in the space of a representation which form an orbit with
respect to the Weyl group.

All vectors in the space of a representation can be numbered by the
Gel'fand-Zetlin patterns:
\eqn\RIS{\eqalign{
m_{1,N}\, m_{2,N} \quad \ldots \quad \quad\ldots \quad\quad m_{N-1,N}\,
m_{N,N}\cr
\qquad m_{1,N-1}\, m_{2,N-1} \ldots \ldots  m_{N-2,N-1}\, m_{N-1,N-1} \quad \cr
\ldots \qquad \qquad \ldots \cr
m_{13}\, m_{23}\, m_{33} \cr m_{12}\, m_{22} \cr m_{11}
}}
The rows in eq. (A9) are the highest weights in the following sequence
of embeddings [19]
\eqn\BEGIN{
{\rm U(N)} \supset  {\rm U(N-1)} \supset   \ldots
\quad \ldots  \supset  {\rm U(2)} \supset  {\rm U(1)}   }
where U($k$) is spanned by the first $k$ rows and columns of the U(N)
matrix in the fundamental representation. The first row in eq. (A9)
coincides with the U(N) highest weight and is fixed. Other numbers are
restricted by the single condition : $m_{k,l} \ge m_{k,l-1} \ge
m_{k+1,l}$. The character of U(1) is equal to $\exp (im_{11}\varphi
_1)$ and, hence, in order to get a non-zero answer in eq. (A6) one has
to take $m_{11} = 0$. From the identity U($k$)=U(1)$\times$SU($k$), it
follows that $\hchi _{u(k)}(e^{i\varphi}) =
\hchi_{su(k)}(e^{i\varphi})
\exp i\sum _{j=1}^k m_{j,k}(\varphi_1 + \varphi_2 + \ldots+ \varphi_k)$
and one has to take the trivial representation for all such U(1)
subgroups:
\eqn\RK{
r_k = \sum _{j=1}^k m_{n,k} = 0    }
Eq (A11) has to be fulfilled for all $k \le N$ and corresponding
vectors will have the zero weight (by definition, the weight of a
vector is equal to $[r_1,r_2 - r_1,\ldots ,r_N - r_{N-1}]$). When
$k=N$, eq.  (A11) gives the selection rule (3.14).

\appendix{B}{}

Using the decomposition (3.1) and the eqs.(3.2), (3.3), we can
represent (4.23) (in the functional integral formalism (2.1),(2.2),
but with free boundary conditions for $\varphi (0)$ and $\varphi
(\beta )$) as
\eqn\ALPHA{\eqalign{
 K_{\alpha _1\alpha _2}(\beta ) = \sum^N_{m,n=1} \langle e^{\alpha
 _1z_m(0)}\vert
\big( \Omega (0)\Omega ^+(\beta )\big) _{mn} \vert ^2e^{\alpha
 _2z_n(\beta )}\rangle
}}
     
Note that, for the free boundary conditions, the constraint (3.5) is
absent due to the linear measure for the hermitian matrix field,
$A(t)$, with the Gaussian weight (see eq.(3.2)).  This functional
integral is ultra-local in time, and one has only to take care of time
ordering.  The Vandermonde determinants of the z-variables cancel
one another after the integration over $A(t)$ such that only two of
them remain at the end points (see [1])
     
As the result we get
\eqn\KAL{\eqalign{
& K_{\alpha _1\alpha _2}(\beta ) = \sum ^N_{m,n=1} \int \prod ^N_{k=1}
\big[ d z_k(t) e^{-{1\over N}tr\int ^\infty _{-\infty }dt({1 \over 2}{\dot
 z}_k^2 +
N^2V(z_k))}\big] \Delta (z(-\infty ))\Delta (z(\infty ))  \cr
& \times e^{\alpha _1z_m(0) + \alpha _2z_n(\beta )}\ \bigg(
 {\rm Texp}\big\{ -{1 \over 2N} \int ^\beta _0 dt\, \widehat Q(z)\big\}
\bigg)^{mn}_{mn}
}}
where the matrix $\widehat Q(z)$ is of the form
\eqn\WIDE{
\widehat Q(z)^{mn}_{ij} =
\delta^{mn} \delta_{ij}
\bigg(\sum_{k\ne i}{1\over (z_i - z_k)^2} +
 \sum_{k\ne n}{1\over (z_k - z_n)^2} \bigg) - 2{\delta^m_i \delta_j^n
\over (z_i - z_j)^2} }
where the double indices $(i,m)$ and $(j,n)$ represent ones in the
space of the adjoint representations. The unitarity condition,
$\Omega\Omega ^+ =I$, leads to the equations
\eqn\SUM{
\sum_{i=1}^N \widehat Q_{ij}^{in} = 
\sum_{j=1}^N \widehat Q_{ij}^{mj} = 0,   }
 hence the number of independent indices is equal to the dimension of
the adjoint, $d_{\rm adj} = N^2 -1$.
     
In the Hamiltonian language eqs. (B2), (B3) are equivalent to eqs.
(4.26), (4.27) (where it was already taken $V(z) = -{z^2 \over 2}$).
     
\appendix{C}{}
     
One has to calculate the sum over ordered integer numbers of the type
\eqn\XIRQ{ \eqalign{
\Xi_R(q) = \sum_{n_i \geq n_{i+1}} M^S_{R\otimes S} \quad q^{\sum^N_{i=1}n_i}
}}
where $n_i$ are the components of the highest weight of a
representation S. The coefficients $M^S_{R\otimes S}$ can be found by
the following graphical rules [21]. Let us draw the vertical line and
place the Young tableau of a representation $R$ in such a way that the
number of boxes to the right of the line equals to the empty space to
the left. After that one has to add all boxes of the tableau of a
representation $S$ to the first tableau ($R$) so that to reproduce the
second one ($S$) to the right of the line according to the following
algorithm. One has to place all boxes from the first row of the second
tableau in different columns of the first one so that they have formed
a possible Young tableau. Then one has to proceed with boxes from the
second row but with the restriction that above and to the right of
each box in an obtained tableau the number of them must not exceed the
number of boxes from the first row. Then one has to repeat this
procedure for all rows.  It is convenient to attach labels to all
boxes of the second tableau, boxes from the same row having the same
labels. Then $M^S_{R\otimes S}$ will be equal to the number of all
possible combinations of labels.  The structure of labeling is not
fixed only in columns in which there are boxes of the first Young
tableau. Therefore, one can consider every labeling in these columns
as a separate sum over $n_i$-s contributing to eq.(C1) which takes in
this case the form
\eqn\XIR{
\Xi _R(q) = \sum_{\{L\}} \sum_{\{n_i\}} q^{\sum^N_{i=1}n_i}
}
where $\sum_{\{L\}}$ is the sum over all possible labelings;
 $\sum_{\{n_i\}}$ is the sum over N integer numbers corresponding to
 all possible tableaux compatible with a particular labeling from the
 first sum.  All numbers run to the infinity but from below the sum is
 restricted by a condition more complicated than the simple ordering.
 Nevertheless this sum gives just
\eqn\QALX{
q^{a(L)}\Xi _{\rm sing}(q)
}
with an integer $a(L)$, hence eq.(4.50) is true and
\eqn\PRQ{
P_R(q) = \sum_{\{ L\} }q^{a(L)}   }
\eqn\ALS{   a(L) = \sum^N_{k=1}k\Delta _k   }
where $\Delta _k = min(n_k - n_{k+1})$, $ (n_{N+1}=0)$ are minimal
differences between the indices in the second sum in eq.(C2).
     
If all labels less than k have been already placed in some way, one
 has a configuration which is equivalent to the case when the number
 of rows is equal to $N-k$ and placing k-labels one gets a number of
 configurations with the number of rows less by 1.  It gives rise to
 recurrence relations between contributions to $P_R(q)$ from different
 tables with the decreasing number of rows.  For example, for the
 sequence of representations defined by eq. (4.52) there are only two
 ways to place $k$-labels:

\bigskip
    \vskip 5pt
    
    %
\hskip 10pt
\epsfbox{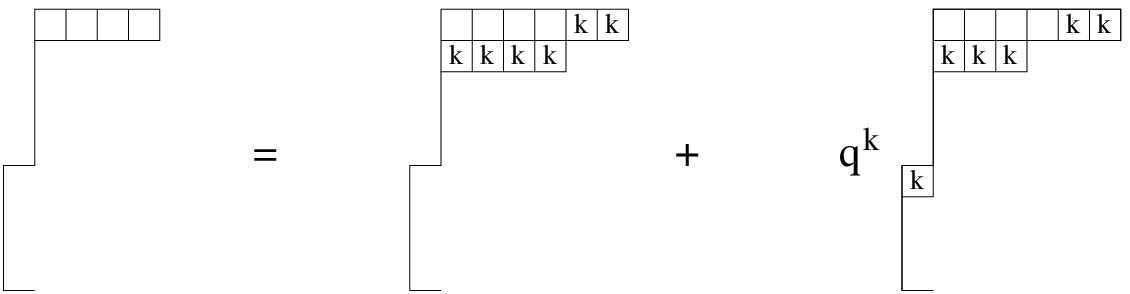}
\vskip -10pt
\centerline{}
 
    \bigskip

\eqn\BEGB{
a_k^{(n)} = a_{k+1}^{(n)} + q^k a_{k+1}^{(n-1)}  }
and $P_{A_n}(q) = a_1^{(n)}$. Eq. (4.54) is a direct consequence of
eq.  (C6). For representations (4.53), the recurrence relation is a
little bit more complicated:

\bigskip
    \vskip -40pt
    
    %
\hskip -70pt
\epsfbox{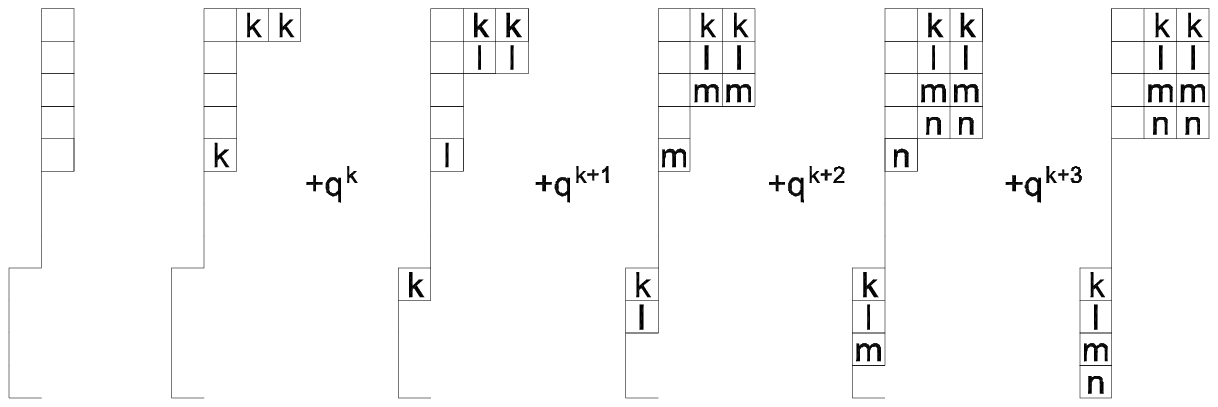}
\vskip -120pt
\centerline{}
 
    \bigskip

\eqn\PIKT{
b_k^{(n)} = \sum _{i=0}^{n} q^{ik} b_{k+1}^{(n-i)}
}
and $P_{B_n}(q) = b^{(n)}_1(q)$.  It is clear that eq. (C6) describes
fermions (no more than one particle can fill the $k$'th level) and
eq. (C7) describes bosons (the number of particles are not
restricted). One can proceeds in the same way for other
representations but the structure of answers will be the more
complicated the more non-trivial Young tableaux are. Nevertheless,
they can be always interpreted as statistical sums for $n$ particles
in the equidistant level system, where $n$ is the number of positive
(or negative) components of a highest weight. For example, for the
sequence of representations
\eqn\CNN{
C_n = (\underbrace{1,\ldots,1}_{n},0,\ldots,0,\underbrace{-1,\ldots,-1}_{n})
}
one gets the recurrence relation:

\bigskip
    \vskip -120pt
    
    %
\hskip -90pt
\epsfbox{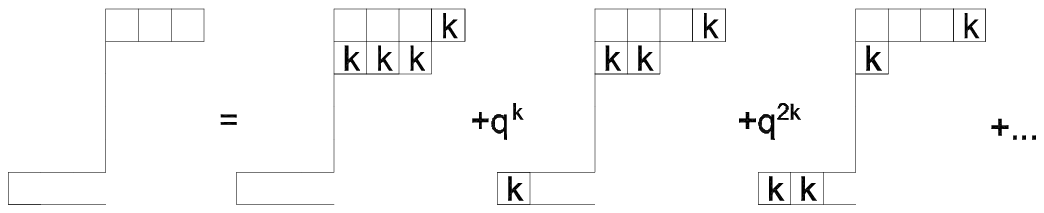}
\vskip -120pt
\centerline{}
 
    \bigskip

\eqn\KKKK{
c_k^{(n)} = c_{k+1}^{(n)} + \sum _{i=1}^{n} q^{k+i-1} c_{k+i+1}^{(n-i)}
}
which describes fermions but with "wrong" weights: the weight of a
fermion on the $k$-th level is equal to $q^{k-j}$ where $j = 0$ if the
$(k-1)$-th level is empty and $j = k - 1$ otherwise.  The form of the
polynomials for several simplest representations are: i) for the
adjoint
\eqn\ADJ{
P_{adj}(q) =  {q - q^N \over 1 - q}
}
ii) for 4 representation with $n=2$ \foot{the representation
conjugate to $A_2$ gives the same answer.}
\eqn\ARR{\eqalign{
P_{A_2} &= {q^3(1-q^{N-2})(1-q^{N-1}) \over (1-q)(1-q^2)} \cr
P_{B_2} &= {q^2(1-q^{N-1})(1-q^N) \over (1-q)(1-q^2)} \cr
P_{C_2} &= {q^2(1-q^{N-3})(1-q^N) \over (1-q)(1-q^2)}
}}

\centerline{\tit \bf References}
     
\parskip 4pt
\parindent 35pt
\hsize=5.8in

\hskip -48pt 1.
{\narrower
 V.A.Kazakov, preprint LPTENS 90/30, December 1990;
\smallskip}

\hskip -53pt 2.
{\narrower
D.Gross and I.Klebanov, Nucl.Phys. {\bf B344} (1990) 475;
\smallskip}

\hskip -53pt 3.
{\narrower
D.Gross and I.Klebanov, Nucl.Phys. {\bf B354} (1990) 459;
\smallskip}

\hskip -53pt 4.
{\narrower
Marchesini and Onofri, J.Math.Phys. {\bf 21} (1980) 1103;
\smallskip}

\hskip -48pt 5.
{\narrower E.Brezin, C.Itzykson, G.Parisi and J.-B.Zuber,
Comm.Math.Phys. {\bf 59} (1978) 35;
\smallskip}

\hskip -48pt 6.
{\narrower
V.A.Kazakov and A.A.Migdal, Nucl.Phus. {\bf B311} (1988) 171;
\smallskip}

\hskip -48pt 7.
{\narrower E.Brezin, V.A.Kazakov and Al.B.Zamolodchikov,
Nucl.Phys. {\bf B338} (1990) 673;
\smallskip}

\hskip -48pt 8.
{\narrower
D.Gross and N.Miljkovic, Phys.Lett. {\bf B238} (1990) 217;
\smallskip}

\hskip -48pt 9.
{\narrower
P.Ginsparg and J.Zinn-Justin, Phys.Lett. {\bf B240} (1990) 333;
\smallskip}

\hskip -48pt 10.
{\narrower
J.Parisi, Phys.Lett. {\bf B238} (1990) 209, 213;
\smallskip}

\hskip -48pt 11.
{\narrower
G.Moore, preprint RU-91-12 (1991), Nucl.Phys. {\bf B368} (1992) 557;
\smallskip}

\hskip -48pt 12.
{\narrower
I.Kostov, Phys.Lett. {\bf B215} (1988) 499;
\smallskip}

\hskip -53pt 13.
{\narrower
V.L.Berezinski, JETP {\bf 34} (1972) 610;
\smallskip}

\hskip -53pt 14.
{\narrower
J.M.Kosterlitz and D.J.Thouless, J.Phys. {\bf C6} (1973) 1181;
\smallskip}

\hskip -53pt 15.
{\narrower
J.Villain, J.Phys. {\bf C36} (1975) 581;
\smallskip}

\hskip -53pt 16.
{\narrower
C.Itzikson and J.B.Zuber, J.Math.Phys. {\bf 21} (1980) 411;
\smallskip}

\hskip -53pt 17.
{\narrower
D.P.Zelobenko, "Compact Lie Groups and Their
Applications",\hfill \break
Nauka 1970,Moscow (in Russian);\hfill \break
 Amer.Math.Soc.Translations, {\bf 40}
(1973);
\smallskip}

\hskip -53pt 18.
{\narrower
 Y.Nambu, "Fermions living in a space of Lie group", Enrico Fermi Inst. 
preprint (1984);
\smallskip}

\hskip -53pt 19.
{\narrower
A.N.Kirillov, J.Geom.and Phys. {\bf 5} (1988) 365;
\smallskip}

\hskip -53pt 20.
{\narrower
A.O.Barut and R.Raczka, "Theory of Group Representations and
Applications", 
World Scientific 1986,Singapore, Ch. 10, \S 1;
\smallskip}

\hskip -53pt 21.
{\narrower
 ibid, Ch. 8, \S 8;
\smallskip}
     
\hskip -53pt 22.
{\narrower
S.Das and A.Jevicki, Mod.Phys.Lett. {\bf A5} (1990) 1639.
\smallskip}
     
\vfill
\eject     
     
\bye